\documentclass[superscriptaddress,nofootinbib,twocolumn,prb]{revtex4}

\usepackage{amsmath,amssymb}
\usepackage{graphicx}
\usepackage{color}
\usepackage{hyperref}
\usepackage{url}
\usepackage[utf8]{inputenc}
\usepackage{slashed}
\usepackage{subfigure}
\usepackage{slashed,bbm}
\usepackage{graphics,psfrag,epsfig}
\usepackage{dsfont}
\usepackage{setspace}

\newcommand{\abs}[1]{ \left|  #1 \right| }

\allowdisplaybreaks

\begin{document}

\title{Fluctuation-induced continuous transition and quantum criticality in Dirac semimetals}

\author{Laura~Classen}
\affiliation{Condensed Matter Physics and Materials Science Division, Brookhaven National Laboratory, Upton, NY 11973-5000, USA}

\author{Igor~F.~Herbut}
\affiliation{Department of Physics, Simon Fraser University, Burnaby, Canada}

\author{Michael~M.~Scherer}
\affiliation{Institute for Theoretical Physics, University of Cologne, D-50937 Cologne, Germany}

\begin{abstract}
We establish a scenario where fluctuations of new degrees of freedom at a quantum phase transition change the nature of a transition beyond the standard Landau-Ginzburg paradigm. To this end we study the quantum phase transition of gapless Dirac fermions coupled to a $\mathbb{Z}_3$ symmetric order parameter within a Gross-Neveu-Yukawa model in 2+1~dimensions, appropriate for the Kekul\'e transition in honeycomb lattice materials.
For this model the standard Landau-Ginzburg approach suggests a first order transition due to the symmetry-allowed cubic terms in the action.
At zero temperature, however, quantum fluctuations of the massless Dirac fermions have to be included. We show that they reduce the putative first-order character of the transition and can even render it continuous, depending on the number of Dirac fermions $N_f$. A non-perturbative functional renormalization group approach is employed to investigate the phase transition for a wide range of fermion numbers. For the first time we obtain the critical $N_f$, where the nature of the transition changes. Furthermore, it is shown that for large $N_f$ the change from the first to second order of the transition as a function of dimension occurs exactly in the physical 2+1~dimensions. We compute the critical exponents and predict sizable corrections to scaling for $N_f =2$.
\end{abstract}

\maketitle

\section{Introduction}

\subsection{Quantum Phase Transitions}

Correlated many-body systems display a variety of complex collective phenomena that are related to the appearance of phase transitions.
In particular, quantum phase transitions (QPT)\cite{sachdev2011}, i.e.~zero-temperature phase transitions driven by quantum fluctuations, are believed to be key to the understanding of unconventional properties of correlated many-body systems, even at finite temperature.
For example QPTs are considered to play an important role in the phase diagram of high-temperature superconductors~\cite{sachdev2003}.
In general, the established theoretical framework for the description of phase transitions is based on the identification of an appropriate order parameter $\phi$ distinguishing a disordered phase with expectation value $\langle\phi\rangle=0$, from a phase with a finite symmetry-breaking order $\langle\phi\rangle >0$, as a function of some tuning parameter\cite{landau2013statistical}.
In the case of a QPT, the tuning parameter is a quantity which controls the strength of the quantum fluctuations, e.g., an interaction strength, a doping level or -- more formally -- the value of Planck's constant.

In a phase transition of second order, the expectation value of the order parameter $\langle\phi\rangle$ is a continuous function of the tuning parameter, and the transition point is referred to as a quantum critical point (QCP). In the vicinity of such a QCP an universal critical behavior emerges.
In many cases, the critical behavior can successfully be accessed by Landau-Ginzburg-Wilson (LGW) theory\cite{Wegner:1972ih,Wilson:1973jj}. It describes the phase transition by a continuum field theory formulated exclusively in terms of the order parameter in the appropriate space-time and restricted by the symmetries of the underlying system. The overwhelming success of this theoretical paradigm notwithstanding, it appears that there are QPTs which demand concepts beyond the LGW paradigm\cite{senthil2004}, deconfined quantum critical points\cite{Senthil1490} being a prime example. These should describe a second-order QPT between two distinct symmetry-broken phases, in contrast to the LGW theory that would suggest either a first order transition or a coexistence region, with both order parameters having a finite expectation value. The reason why the LGW theory could fail is that the deconfined QCP cannot be described purely in terms of order parameter fluctuations. Instead, new degrees of freedom such as spinons emerge right at the critical point, and a more natural formulation of the transition is provided directly in terms of these.
%

\begin{figure}[t!]
\centering
 \includegraphics[width=\columnwidth]{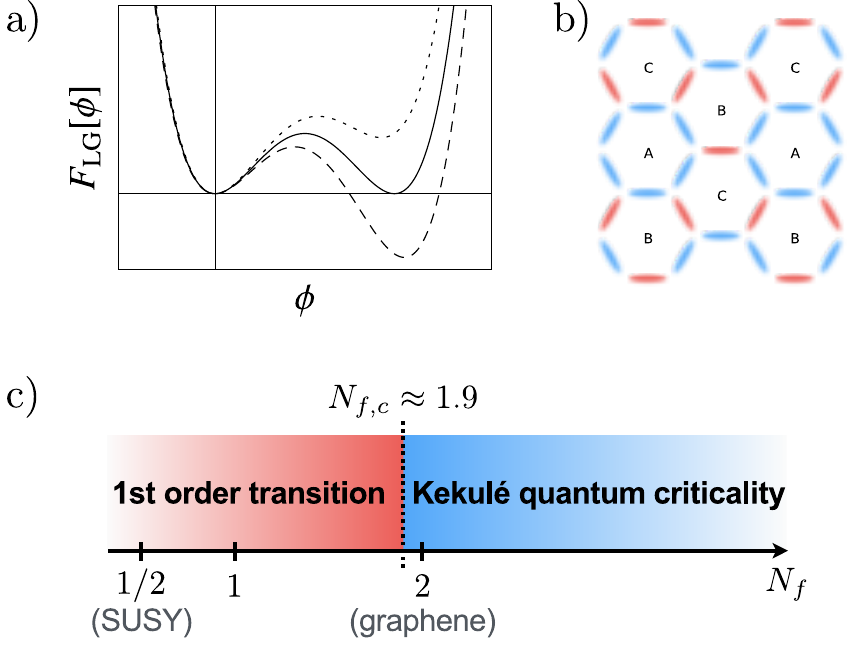}
 \quad\quad
 \caption{(a) Sketch of the LG free energy $F_{\text{LG}}[\phi]=F_0+r\, |\phi|^2 + g (\phi^3+\phi^{\ast 3})+\lambda |\phi|^4$ across the phase transition, where the dotted (dashed)
line corresponds to the unordered (ordered) phase.
(b) Sketch of the Kekul\'e valence bond solid state on the honeycomb lattice. Darker and lighter red bonds mark different hopping amplitudes. The pattern leads to a tripled unit cell and reduced rotational symmetry in the Kekul\'e phase.
(c) Schematic phase diagram for the quantum semi-metal to Kekul\'e VBS transition in two-dimensional Dirac semi-metals. With the functional RG approach, we determine the number of critical fermion flavors for fermion-induced quantum criticality to occur at $N_{f,c}\approx 1.9$. We conclude that the Kekul\'e quantum transition in graphene can be expected to be continuous and give rise to quantum critical behavior.}
\label{fig:kekule01}
\label{fig:schematic}
\end{figure}

\subsection{Fluctuation-induced critical points}

Similarly to the emergent spinons at a deconfined QCP, gapless fermions can provide additional degrees of freedom specific to a given quantum phase transition,  so that it becomes necessary to go beyond a description purely in terms of the order parameter fields.
For example, the presence of gapless fermion fluctuations in continuous quantum phase transitions towards an order parameter with $O(N)$ symmetry severely modifies the quantitative estimates for its critical exponents and defines the fermionic or chiral universality classes\cite{Rosenstein:1992,Rosenstein:1993zf}.
Moreover, the impact of the fermion fluctuations could be so strong that a transition becomes second order, although the pure LGW theory would predict a first order transition. In this case we speak of a fermion-induced quantum critical point\cite{li2015}.
More generally, a fluctuation-induced critical point occurs, when quantum or thermal fluctuations render a putatively first-order transition continuous.
Evidence for a fermion-induced second order transition has been put forward in Refs.~\onlinecite{li2015,Scherer:2016zwz,2016arXiv161007603J} in the context of the 2+1 dimensional honeycomb electrons near the transition to a Kekul\'e valence bond solid (VBS).
A closely related scenario was also proposed in 3D double-Weyl semimetals for the transition to a nodal-nematic order\cite{2016arXiv160906313J}.
The formation of a Kekul\'e VBS has been observed in artificial graphene \cite{gomes2012} and graphene on a Copper substrate \cite{Gutierrez2016}. Theoretically,
the Kekul\'e VBS and the nodal-nematic order are described in terms of
a $\mathbb{Z}_3$-symmetric complex-valued order parameter field $\phi$.
As a consequence of the discrete $\mathbb{Z}_3$ symmetry,
a cubic term $\propto (\phi^3+\phi^{*3})$ is allowed in the LG free energy $F_{\text{LG}}$.
If such a term is present,
the minimum of $F_{\text{LG}}$ jumps discontinuously from $\phi=0$ to $\phi\neq0$ at the phase transition, see Fig.~\ref{fig:kekule01}.
That is this Landau-type mean-field criterion suggests to expect a first-order phase transition in the considered system\cite{PhysRevB.8.3419}.
Indeed, for the three-state Potts model, which exhibits such a $\mathbb{Z}_3$ symmetry, the first-order behavior of the thermal phase transition has been corroborated for three and higher spatial dimensions\cite{RevModPhys.54.235}.
On the other hand, an exact result for the three-state Potts model in two dimensions suggests a second-order phase transition in contrast to the argument above\cite{baxter1973}. In this case the strength of fluctuations is increased by decreasing the dimensionality of the system.

In the case of the Kekul\'e VBS and the nodal-nematic order in Dirac materials, the formulation in terms of a $\mathbb{Z}_3$ order parameter field is insufficient. The strength of fluctuations is increased by the inclusion of additional critical degrees of freedom, such as gapless Dirac fermions.
In fact, it is argued in Refs.~\onlinecite{li2015,Scherer:2016zwz} (Kekul\'e) and Ref.~\onlinecite{2016arXiv160906313J} (nodal-nematic) that the first-order behavior breaks down and a critical point is induced if the number of fermions $N_f$, corresponding to the number of pairs of Dirac points, is sufficiently large (cf. also Ref.~\onlinecite{Rosenstein:1992}).
That is, there should be a critical value $N_{f,c}$ so that the transition is of first order for $N_f<N_{f,c}$ and of second order for $N_f>N_{f,c}$.
However, the precise value of the critical fermion number is presently not known.
Moreover, so far there is no non-perturbative calculation which would access both the first and the second order regimes.
In Ref.~\onlinecite{2016arXiv161007603J} Majorana quantum Monte Carlo simulations have been performed for $N_f\in\{2,\ldots,6\}$, which all yield a second order transition. On the other hand, Refs.~\onlinecite{Scherer:2016zwz} and \onlinecite{2016arXiv160906313J} used perturbative expansions around the upper critical dimension and large $N_f$ to study the fermion-induced critical point.

The question whether fluctuations can indeed alter the nature of a phase transition relative to the LGW prediction is inherently non-perturbative, as it requires sizable fluctuation effects. Thus a non-perturbative study which would access the full range of $N_f$ is required in order to establish the possible occurrence of a fermion-induced QCP. Here we therefore perform such a study through an investigation of the zero-temperature Kekul\'e transition in 2+1 dimensional Dirac materials with the help of the non-perturbative functional renormalization group (FRG)\cite{wetterich1993}.
The FRG has been proven to be a suitable tool in the study of non-perturbative aspects of quantum and statistical field theories, see, e.g., Refs.~\onlinecite{Berges:2000ew,Polonyi:2001se,Pawlowski:2005xe,Gies:2006wv,Delamotte:2007pf,Braun:2011pp,metzner2011} for reviews and Sec.~\ref{sec:frg} for important aspects related to this work. Once the appearance of a fermion-induced QCP is established, the question arises if the vicinity to a first order transition affects the critical behavior in the second order transition regime close to $N_{f,c}$. We therefore compute the critical exponents characterizing the second order transition, and determine and discuss the corrections to scaling.

\subsection{Main results}

In this work we study whether strong fluctuations can induce a second order phase transition and consequently critical behavior.
To that end, we consider the Kekul\'e ordering transition at zero temperature in two-dimensional Dirac semimetals, where the conventional Landau-Ginzburg paradigm of phase transitions would predict a discontinuous transition.
By means of the non-perturbative functional renormalization group, we investigate a full range of the number of Dirac fermions $N_f$ for this model.
We indeed find that fluctuations render the phase transition continuous in $D=2+1$ dimensions if $N_f$ is sufficiently large.
This is sketched in Fig.~\ref{fig:schematic}. We also obtain the first direct estimate for the critical fermion flavor number, where the nature of the transition changes.
We find $N_{f,c}\approx1.9$, which lies in the limits provided by quantum Monte Carlo ($N_f<2$) and SUSY ($N_{f,c}>1/2$) calculations.
In particular, spin-1/2 fermions on the honeycomb lattice, $N_f=2$, show a second order transition with quantum critical behavior.
In the critical regime $N_f>1.9$, we determine the critical exponents characterizing the scaling of correlation functions.
We find the second largest critical exponent to be negative but small in magnitude, which implies a nearly marginal RG direction in the theory.
In the limit of a large number of Dirac fermions, the corresponding coupling becomes exactly marginal.
This leads to sizable corrections to scaling close to the quantum critical point for the whole $N_f$-regime.
We expect these corrections to be detectable by QMC simulations.

\subsection{Outline}

We use a Gross-Neveu-Yukawa theory to model the quantum phase transition of massless Dirac fermions towards the Kekul\'e valence bond solid (VBS), which we motivate in Sec.~\ref{sec:model}.
The Kekul\'e VBS is described in terms of a complex order parameter field with $\mathbb{Z}_3$ symmetry, Yukawa-coupled to Dirac fermions.
In Sec.~\ref{sec:frg}, we explain the renormalization group picture of phase transitions and introduce renormalization group fixed points and critical exponents. This allows a more specific notion about what sizable fluctuations mean in the considered context.
We also present our functional RG approach and discuss its key features.
In Sec.~\ref{sec:results}, we discuss our results for the fermion-induced quantum critical point.
We carefully establish convergence within our truncation scheme and compare to previous studies if possible.
In particular, we also highlight the advantages of the approach presented here. We give our best estimate for the fermion number, where the nature of the transition changes, and determine the critical behavior in the second order regime.
Finally, we draw conclusions in Sec.~\ref{sec:conc}. Technical details are relegated to the appendices.

\section{Semimetal to Kekul\'e VBS transition}\label{sec:model}

\subsection{Honeycomb fermions and Dirac semimetals}

Since the discovery of graphene, considerable interest has been generated by Dirac materials, that impact a broad range of research from fundamental physics to concrete technological applications\cite{vafek2013, wehling2014}.
More specifically, charge-neutral graphene\cite{Novoselov2005,Geim2007,castroneto2009} serves as a the prototypical example for a class of materials where low-energy Dirac excitations emerge as a result from the underlying honeycomb lattice structure.
A wide variety of possible ordering patterns have been proposed to become relevant to the phase diagram of interacting Dirac electrons on the honeycomb lattice\cite{sorella1992,herbut2006,hou2007,honerkamp2008,raghu2008,herbut2009,ryu2009,roy2010c, weeks2010,classen2014,volpez2016,sanchez2016}.
Among these ordered states, the Kekul\'e valence bond solid (VBS)~\cite{hou2007,ryu2009,roy2010c,classen2014}  --~a particular dimerization pattern of the fermions on the honeycomb lattice~-- has recently attracted special attention due to its experimental realization~\cite{gomes2012,Gutierrez2016}.

Explicitly, we consider two-dimensional materials whose low-energy effective theory can be expressed in terms of a free Dirac Lagrangian\cite{Semenoff1984}
\begin{align}
\mathcal{L}_{\psi}=\bar\psi\,\gamma_\mu\partial_\mu\psi\,,
\end{align}
where the conjugate field is defined as $\bar\psi=\psi^\dagger\gamma_0$ and  $\partial_\mu=(\partial_\tau, \vec{\nabla})$ denotes the imaginary-time and space derivative.
We have set the Fermi velocity to unity, $v_F=1$.
The $\gamma$~matrices satisfy the Clifford algebra $\{\gamma_\mu,\gamma_\nu\}=2\delta_{\mu\nu}$ and their four-dimensional representation in $D=2+1$ dimensions reads
\begin{align}
\gamma_0=\mathds{1}_2\otimes\sigma_z,\quad \gamma_1=\sigma_z\otimes\sigma_y,\quad\gamma_2=\mathds{1}_2\otimes\sigma_x\,,
\end{align}
with the Pauli matrices $\sigma_i$.
In momentum space, we have $\psi(x)=\int d^Dqe^{iqx}\psi(q)$
with momentum vectors $q=(\omega,\vec{q})$ gathering Matsubara frequency $\omega$ and wavevector $\vec{q}$.
The four spinor components of the Dirac field $\psi$ can be related to the two sublattice and the $K,-K$ valley degrees of freedom of spinless fermions on the honeycomb lattice.
Furthermore we define
\begin{align}
\gamma_3=\sigma_x\otimes\sigma_y,\quad \gamma_5=\sigma_y\otimes\sigma_y\,,
\end{align}
which anticommute with all $\gamma_\mu$. Their combination given by $\gamma_{35}=-i\gamma_3\gamma_5$ commutes with all $\gamma_\mu$ and anticommutes with $\gamma_3$ and $\gamma_5$.
As a convenient generalization of this model, we introduce the number of fermion flavors of four-component spinors by $\psi^\dagger\rightarrow\psi_i^\dagger$
with $i\in\{1,\ldots,N_f\}$.
This corresponds to an arbitrary number of flavors or of pairs of Dirac points $\pm\vec{K}_i$ in the fermion spectrum.
$N_f=2$ then corresponds to an eight-component spinor, which describes the ``graphene case'' of spin-1/2 fermions on the honeycomb lattice with spin-rotation invariance.
%

\subsection{Kekul\'e valence bond solid}\label{sec:modelsub}

We are interested in the phase transition from the above Dirac-semimetal phase into the Kekul\'e valence bond solid\cite{hou2007}.
A sketch of this state on the honeycomb lattice is shown in Fig.~\ref{fig:kekule01}, exhibiting varying bond strengths between nearest-neighbors.
In other words, the uniform nearest-neighbor hopping amplitude of the symmetric phase receives two different, but real values in the symmetry-broken phase, arranged in the specific pattern shown in the figure.
This pattern breaks the translational symmetry of the original lattice and reduces the sixfold rotation symmetry to a threefold one.
In the low-energy effective theory, the modulation of the nearest-neighbor hopping is described by a complex order parameter $\phi=(\phi_1+i\phi_2)/\sqrt{2}$, whose phase controls the angles of the dimerization pattern\cite{ryu2009} and in our case has to be $\mathbb{Z}_3$ symmetric.
This is related to the presence of three degenerate Kekul\'e ground states as obtained by translation of the pattern shown in Fig.~\ref{fig:kekule01} along the lattice vectors.
The $\mathbb{Z}_3$ symmetric order parameter couples to the fermions via a Yukawa term of the form \cite{roy2010c}
\begin{align}
\mathcal{L}_{\psi\phi}=i \bar h\bar\psi\left(\phi_1 \gamma_3 +\phi_2 \gamma_5\right)\psi\,.
\end{align}
We note that this part of the action is still invariant under the general ``chiral'' continuous $U(1)$ transformation
\begin{align}\label{eq:chiralU1}
\psi\rightarrow e^{i\theta \gamma_{35}/2}\psi,\quad \bar\psi\rightarrow \bar\psi e^{-i\theta \gamma_{35}/2}\,,\quad
\phi\rightarrow e^{i\theta}\phi\,,
\end{align}
corresponding to translations on the honeycomb lattice.
This property will be violated by the bosonic action
\begin{align}\label{eq:bos}
\mathcal{L}_\phi&=-\phi^*\partial_\mu^2\phi +\bar m^2\abs{\phi}^2 + \bar g(\phi^3+\phi^{*3}) + \bar\lambda \abs{\phi}^4 \\[5pt]
& +\bar g_5(\phi^3+\phi^{*3})\abs{\phi}^2 + \bar g_6(\phi^3+\phi^{*3})^2 +\bar \lambda_6\abs{\phi}^6 +\ldots\notag
\end{align}
in which the terms $\propto(\phi^3+\phi^{*3})$ are responsible for the reduction of the continuous $U(1)$ symmetry down to $\mathbb{Z}_3$.
In general higher order couplings can also occur as denoted by the ellipsis. Here, we have added couplings up to the order $\phi^6$. Dimensional analysis yields the following canonical dimensions for the bosonic couplings
\begin{align}\label{eq:canondim}
[\bar m^2]&=2,\quad [\bar g]=3-D/2,\quad [\bar \lambda]=4-D\,,\\[5pt]
[\bar g_5]&=5-3D/2,\quad [\bar g_6]=[\bar \lambda_6]=6-2D\,,
\end{align}
as well as $[\bar h]=(4-D)/2$ for the Yukawa coupling.
Accordingly, couplings of higher order than $\phi^4$ are omitted in perturbative calculations close to $D=3+1$ dimensions as they become irrelevant (see next section).
In three spacetime dimensions, however, couplings up to $\phi^6$ are relevant, and we will indeed find that they do play a role in deciding the nature of the phase transition.

In equation~\eqref{eq:bos}, we employed the low-energy effective particle-hole symmetry of the model to exclude a linear term in the time-derivative of the order parameter Lagrangian.
Furthermore, we have written down a Lorentz-symmetric kinetic term, i.e. we have set the boson velocity to the same value as the Fermi velocity $v_B=v_F=1$.
This Lorentz-symmetric form of the total action
$S=\int d\tau d^{D-1}x \, ( \mathcal{L}_\psi + \mathcal{L}_{\psi\phi}+\mathcal{L}_\phi)$
however, is not dictated a priori; rather, the
Lorentz symmetry near fermionic quantum critical points with equal velocities for fermions and bosons has been argued to emerge naturally in the deep infrared regime in a large class of Yukawa theories of the same kind\cite{Roy:2015zna},  even if $v_F\neq v_B$ on intermediate scales.
Lorentz symmetry also emerges in our case, as the quantum critical point which we will find is located at vanishing $U(1)$-breaking couplings, i.e.~$\bar g^\ast=\bar g_5^\ast=\bar g_6^\ast=...=0$ and therefore the analysis given in Ref.~\onlinecite{Roy:2015zna} is directly applicable.
Away from the QCP, and in particular in the symmetry broken phase, however, it could also be interesting to consider the case with two different velocities.
We will leave this issue for future work.

\section{Renormalization Group approach}\label{sec:frg}

\subsection{Renormalization group scaling argument}\label{sec:rgscaling}

Before we explain the details of our RG approach,
we would like to translate the picture from the saddle-point approximation of the free energy
into a renormalization group language.
In the RG picture, a system is considered scale-dependent, i.e.~the couplings describing it evolve with the scale when, for example, the energy scale is lowered.
Critical behavior near a second-order phase transition is related to the presence of an infrared attractive renormalization group fixed point of this scale evolution. At a RG fixed point, the system becomes scale invariant by definition.
Consequently, the free energy adopts a scaling form accompanied by universal critical exponents.
In the standard scenario, the critical point is approached by fine-tuning of a parameter, e.g., the temperature $T\to T_c$ for a thermal transition and, e.g., a doping level or a coupling strength for a quantum transition.
In the renormalization group a tuning parameter translates to the presence of an RG relevant direction, which  is infrared repulsive, i.e. the corresponding direction also has to be fine-tuned to approach an otherwise infrared-attractive RG fixed point.
Therefore, in order to find a scaling solution and observe critical behavior in a system, an identical number of physical tuning parameters and RG relevant directions is required.

Dimensional analysis, or power counting, gives a first hint about the RG relevance of the parameters of a model and is valid near the Gau\ss ian fixed point.
At a non-Gaussian fixed point, however, the predictions from dimensional analysis about the RG scaling of parameters receive fluctuation corrections. They can affect the relevance of the RG directions.
A positive (negative) power counting dimension suggests a relevant (irrelevant) parameter.
Parameters with vanishing power counting dimension are called marginal.
In our case, the canonical dimensions for the lowest order couplings appearing in the Lagrangian in Eq.~\eqref{eq:bos} are presented in Eq.~\eqref{eq:canondim}.
Here, $D$ specifies the spacetime dimension of the system.
The power counting dimension of the parameter $m^2$ is strongly relevant and certainly needs to be fine-tuned to approach the critical point.
In contrast, the other power counting dimensions depend on $D$, so the consideration is more subtle:
For example, in $D=4-\epsilon$ dimensions, we observe $[\bar \lambda]=\epsilon$ and the coupling $\bar \lambda$ therefore corresponds to a slightly relevant direction controlled by the size of $\epsilon$.
On the other hand, the critical behavior of this type of continuous field theories below four dimensions is typically governed by non-Gau\ss ian fixed points (NGFP)\cite{wilson1972} and the dimensional counting is modified.
The sign and the magnitude of the fluctuation contribution generally depends on the dimensionality and possibly on the coupling to other degrees of freedom.
When these modifications are large enough, they can change the sign of the power counting dimension of a canonically relevant coupling.
In fact, at the non-Gaussian $O(2)$ Wilson-Fisher fixed point, the $\lambda$ direction is turned irrelevant by a negative contribution from the fluctuations, i.e. $[\lambda]_{\text{NGFP}}<0$, and therefore $\lambda$ automatically approaches its fixed point value in the infrared.
It turns out that this reasoning continues to apply even to the case of three dimensions\cite{zinnjustin1998}, when $\epsilon=1$.

In the $\mathbb{Z}_3$ symmetric scenario, however, we have another term in the free energy, i.e. the cubic order parameter term $\propto \bar g$. It generically introduces an additional RG relevant direction with a larger power-counting dimension, e.g., in $4-\epsilon$ dimensions we have $[\bar g]=1+\epsilon/2$.
Rendering this direction irrelevant to obtain a fluctuation-induced critical point requires a more dominant role of fluctuations. This cannot be controlled by a small $\epsilon$ as
\begin{align}\label{eq:flucs}
	[\bar g]_{\text{NGFP}}=1+\frac{\epsilon}{2}+\text{fluctuations}\gtrless 0\,.
\end{align}
Therefore, this scenario needs the fluctuations to provide a large negative contribution with $|\text{fluctuations}|>3/2$ in the case of $D=2+1$ independent of their origin. This constitutes an intrinsically non-perturbative scenario and motivates us to use the functional renormalization group in the following.

\subsection{Fixed points and critical exponents}\label{sec:frgsub}

More explicitly, the renormalization group theory describes the scale dependence of a physical system by providing $\beta$ functions for the different couplings of a model\cite{herbutbook}.
The $\beta$ functions are differential equations that encode the evolution of the system with respect to the energy (or momentum) scale $k$.
Starting from a ``microscopic'' model for a system at some ultraviolet (UV) cutoff scale $k = \Lambda$, one can then infer the low-energy, or infrared (IR), characteristics in terms of the solution of the $\beta$ functions.
In our case, the UV scale $\Lambda$ corresponds to the scale at which our effective model is valid, e.g., the largest energy scale where a description of the material in terms of Dirac excitations is justified.
This is much smaller than the system's bandwidth and going beyond that energy scale would require an appropriate lattice description.

To formulate the renormalization group approach, we introduce the generalized set of dimensionless couplings for the theory by $\alpha_i$, $i \in \{1,2,\ldots \}$ which will be specified below.
The $\beta$ functions can be written in the form $\partial_t \alpha_i=\beta_i(\alpha_1,\alpha_2,\ldots)$, where the change in scale is written in terms of the renormalization group time $t = \ln (k/\Lambda) \leq 0$.
A renormalization group fixed point~$\alpha^*$ is defined by a simultaneous vanishing of all beta functions of the model,
\begin{align}
\beta_i(\alpha_1^\ast,\alpha_2^\ast,\ldots)=0 \quad \forall\ i.
\end{align}
The critical properties and scaling behavior near such a transition are encoded in the linearized RG flow close to the fixed point~$\alpha^*$,
\begin{align}
\partial_t \alpha_i=&\mathcal{B}_{i,j} (\alpha_j^*-\alpha_j)+ \mathcal O\left((\alpha_j^*-\alpha_j)^2\right)\,,\end{align}
where $\mathcal{B}_{i,j}=(\partial\beta_i/\partial \alpha_j)|_{\alpha=\alpha^*}$ is the {\it stability matrix}.
The eigenvalues $\theta_i$ of $(-\mathcal{B}_{i,j})$ are called the {\it critical exponents} and are universal quantities that characterize the scaling laws at the putative continuous phase transition.
For the following discussion, we introduce an ordering according to the size of the critical exponents, i.e.
\begin{align}
\theta_1 \geq \theta_2 \geq ... \,.
\end{align}
All positive critical exponents $\theta_i$ correspond to RG-relevant directions, i.e., the fixed point repels the flow in that direction.
In turn, negative $\theta_i$ are RG irrelevant and correspond to attractive directions.
Fixed points with no more than one relevant direction (corresponding to no more than one positive critical exponent) can be accessed by tuning a single parameter.
In the present context, the system features a single tuning parameter.
Therefore, a stable infrared-attractive fixed point which gives rise to physically accessible quantum critical behavior is specified by a set of critical exponents where only one eigenvalue is positive, $\theta_1>0$, and all others are negative, $\theta_i <0$ for $i >1$.
Thus, the sign of the second largest critical exponent $\theta_2$ as calculated from the stability matrix decides over the stability of a fixed point.

\subsection{Functional RG approach in a nutshell}

For our study on the quantum transition from the semi-metallic state of two-dimensional Dirac fermions to a Kekul\'e VBS state, we employ the functional renormalization group (FRG) approach\cite{wetterich1993}.
It generalizes Wilson's momentum-shell RG approach by modifying the functional integral representation of the partition function $Z=\int_\Lambda \mathcal{D}\Phi\, \exp(-S[\Phi])$
by a regulator insertion in the microscopic action, i.e. $S\rightarrow S + \int_p\frac{1}{2}\Phi(-p)R_{k}(p)\Phi(p)$.
Here, $\Phi$ represents a collective field variable for all field degrees of freedom of a specific model. In our case it includes both, Dirac fermions as well as complex bosons.
The regulator insertion controls how high energy degrees of freedom are integrated out in the RG procedure and
the common Wilsonian momentum-shell RG integration would correspond to a specific choice of $R_k(p)$.
In general the regulator $R_k(p)$ depends on an IR cutoff scale $k$ and the momentum $p$ of the field configurations that are integrated over in the partition function.
In particular, $R_k(p)$ is designed to suppress the low-momentum fluctuations by choosing $R_k(p)>0$ for $p^2<k^2$.
Further, the regulator function satisfies  $R_k(p)\rightarrow \infty$ for $k\rightarrow \Lambda \rightarrow \infty$ and $R_k(p)\rightarrow 0$ for $k/|p|\rightarrow 0$.
We then define the generating functional for the one-particle irreducible correlation functions, i.e., the scale-dependent effective action $\Gamma_k$ as the (modified) Legendre transform of the Schwinger functional $\ln Z_k$.
By choice of the regulator properties, $\Gamma_k$ then interpolates between the microscopic action $S$ and the quantum effective action  $\Gamma_{k\rightarrow0}$ in the IR.

The central equation of the FRG is the Wetterich equation: an exact functional differential equation which describes the RG evolution of $\Gamma_k$. It provides the explicit interpolation procedure between $S$ and $\Gamma$.
The Wetterich equation reads\cite{wetterich1993}
\begin{align}\label{eqn:Wetterich}
\partial_t\Gamma_k = \frac{1}{2}\text{STr}\left[(\Gamma_k^{(2)}+R_k)^{-1}\partial_t R_k\right]\,,
\end{align}
where $\partial_t=k\partial_k$ and we have introduced the Hessian $\Gamma_k^{(2)}$
\begin{align}
\Big(\Gamma_k^{(2)}\Big)(p,q)=\frac{\overrightarrow{\delta}}{\delta\Phi(-p)^T}\Gamma_k\frac{\overleftarrow{\delta}}{\delta\Phi(q)}\,.
\end{align}
In terms of the exact Wetterich equation, the functional RG approach allows to devise suitable truncations for a practical implementation of Wilson's idea of successively integrating out degrees of freedom in the functional integral formalism.

\subsection{Key features of the FRG}

In general, the FRG provides a unified framework to describe universal and non-universal
properties of quantum and statistical field theories within and beyond the scope of perturbation theory.
That means, given an appropriate model, it can not only access the universal behavior
in the vicinity of a continuous phase transition, but also the system's physical properties away from the transition.
In particular, first-order phase transitions can also be studied\cite{tissier2000,jakubczyk2009,braun2010,schaefer2006,rennecke2014}.

The Wetterich equation has a one-loop structure, resembling the one-loop functional determinant.
Therefore, a proliferation of diagrams as they arise beyond leading order in higher-order loop expansions is avoided with the FRG.
On the other hand, the presence of the Hessian in the denominator of the one-loop expression $\Gamma_k^{(2)}$ encodes higher-order effects within the one-loop structure. For example, the threshold effects introduced by this non-perturbative propagator improve conventional one-loop results for critical exponents while maintaining the simple one-loop form. Furthermore, the FRG can be applied in arbitrary dimension and does not require the identification of a small expansion parameter. Instead, approximations in the FRG approach result from of a particular truncation of the space of operators that generate the RG flow, see Sec.~\ref{sec:frgsub}.
In contrast to perturbative RG approaches where typically highly non-trivial resummation techniques are required in order to obtain quantitative results for critical exponents, the FRG method  already incorporates a specific resummation through the threshold effects\cite{delamotte2005,delamotte2010b,delamotte2010c}.
Moreover, the inclusion of chiral fermions is straightforward within the FRG and has been successfully employed in the context of Yukawa models for Dirac materials\cite{Rosa:2000ju,Hofling:2002hj,Gies:2009da,Janssen:2012pq,mesterhazy2012,janssen2014,Classen:2015mar}.
In contrast to lattice field theory, we employ a continuum formulation here, which means that our results are unaffected by finite-volume or discretization artifacts.
Thus, the FRG constitutes a well-suited tool to investigate interacting RG fixed points, in particular in regimes where the validity of the perturbative RG approach is called into question.
From a more general point of view, results from the perturbative RG, functional RG and numerical methods, as, e.g., (quantum) Monte Carlo calculations, complement each other to provide a more comprehensive understanding of physical systems~\cite{Mihaila:2017ble,otsuka2016}.

\section{Fermion-induced QCP from FRG}\label{sec:results}

\subsection{FRG flow equations}\label{sec:frgsub}

The scale-dependent effective action $\Gamma_k$ generally contains all symmetry-compatible operators of the considered theory.
In practice, the application of Wetterich's equation requires truncating the infinite-dimensional space of RG directions or couplings to a closed and tractable subspace.
Here, we employ a scheme following a derivative expansion which we truncate after the leading order. It is based on the construction of our model in Sec.~\ref{sec:model}, and the ansatz for $\Gamma_k$ explicitly reads
\begin{align}
	\Gamma_k&=\int d^Dx\Big[
	Z_{\psi,k}\bar\psi \gamma_\mu \partial_\mu\psi	+i \bar h_k\bar\psi\left(\phi_1\gamma_3+\phi_2\gamma_5\right)\psi\nonumber\\
	&\hspace{0.9cm}-\frac{1}{2}Z_{\phi,k}(\phi_1\partial_\mu^2\phi_1+\phi_2\partial_\mu^2{\phi}_2)+U_k(\bar \rho,\bar \tau)\Big]\label{eq:Gamma}\,.
\end{align}
with the spacetime dimension $D$. The Dirac spinors $\bar\psi, \psi$ come with a number of $N_f d_\gamma$ spinor components, where $d_\gamma$ denotes the dimension of the Clifford algebra.  $\phi_1$ ($\phi_2$) is the real (imaginary) part of a complex scalar field $\phi=(\phi_1+i \phi_2)/\sqrt{2}$.
In the first and second line, we gather the kinetic terms of the boson and fermion fields which are equipped with a scale-dependent, uniform wave function renormalization $Z_{\psi/\phi, k}$.
Additionally, in the first line, we list the Yukawa coupling $\bar h_k$  which also carries a scale-dependence.
Finally, we have introduced the scale-dependent boson effective potential $U_k$ which includes all symmetry-allowed boson self-couplings. We write it as a function of two $\mathbb{Z}_3$ invariants
\begin{align}
\bar \rho&= \phi^\ast \phi= \frac{1}{2}(\phi_1^2+\phi_2^2)\,, \\
\bar \tau&=\phi^3+\phi^{*3}=\frac{1}{\sqrt{2}}(\phi_1^3-3\phi_1\phi_2^2)\,.
\end{align}
For the determination of the renormalization group fixed point and its properties it is particularly convenient to introduce dimensionless quantities. To that end, we define the dimensionless effective potential and the dimensionless Yukawa coupling as
\begin{align}\label{eq:dimlessU}
	u(\rho, \tau)=k^{-D}U\left(\bar \rho,\bar \tau\right)\,, \quad  h^2=\frac{k^{D-4}}{Z_{\phi,k}Z_{\psi,k}^2}\bar h_k^2\,,
\end{align}
with $\bar \rho=k^{D-2}Z_{\phi,k}^{-1} \rho$ and $\bar \tau=k^{3(D-2)/2}Z_{\phi,k}^{-3/2}\tau$. Dimensionfull quantities are denoted with a bar, dimensionless without.
Further, we introduce the boson and fermion anomalous dimensions
\begin{align}\label{eq:etas}
	\eta_\phi=-\frac{\partial_t Z_{\phi,k}}{Z_{\phi,k}}\,,\quad \eta_\psi=-\frac{\partial_t Z_{\psi,k}}{Z_{\psi,k}}\,.
\end{align}
%
\begin{figure}[t!]
\centering
 \includegraphics[width=\columnwidth]{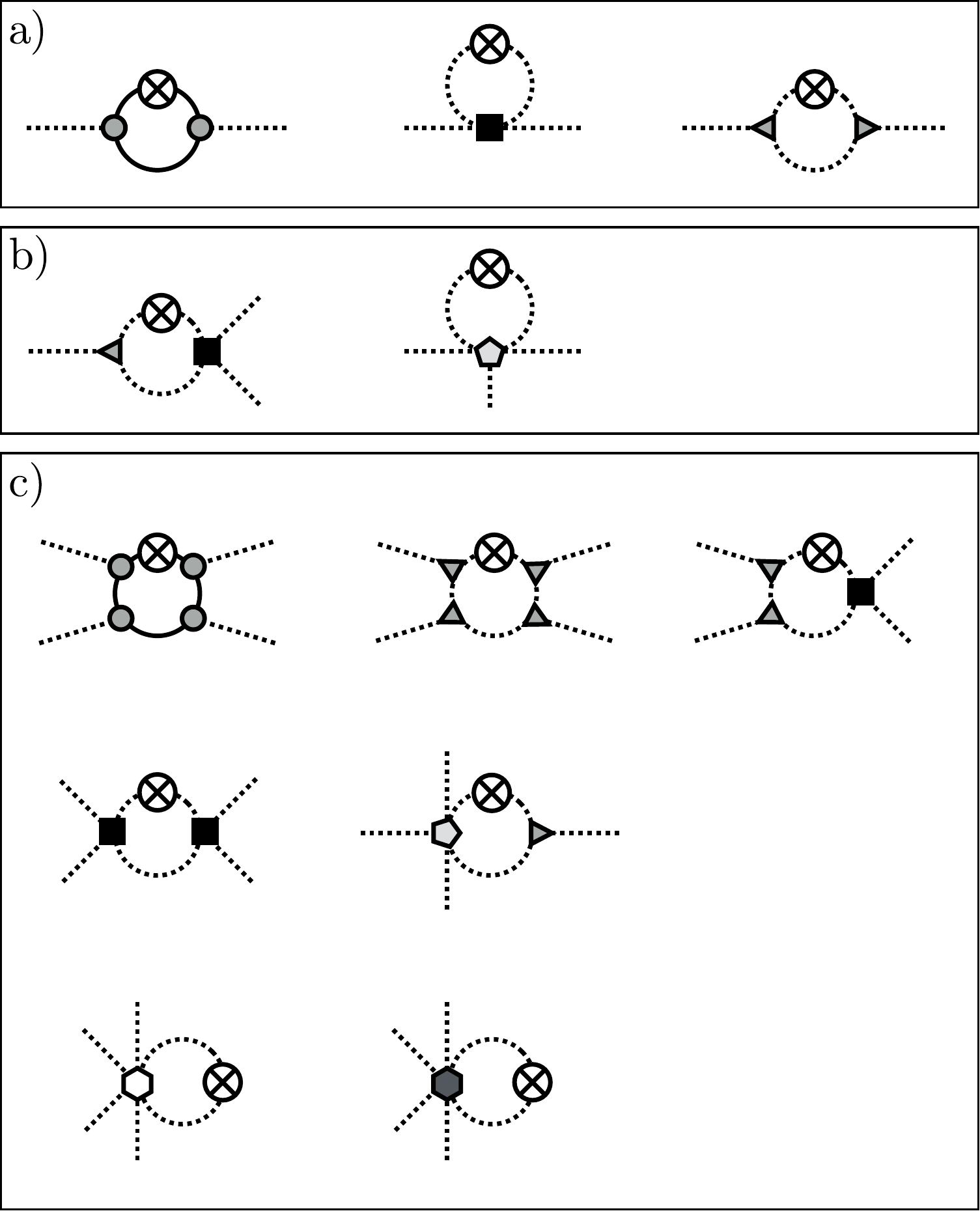}
 \caption{a) Schematic FRG diagrams contributing to the flow of the mass parameter $m^2$.
b) Diagrams contributing to the flow of the cubic coupling $g$.
c) Diagrams contributing to the flow of the quartic coupling $\lambda$. The crossed circle represents the scale derivative of the corresponding propagator. Importantly, the propagator lines are dressed propagators.}
\label{fig:feyn01}
\end{figure}
%
We proceed with the evaluation of the Wetterich equation to calculate explicit expressions for the flow of the effective potential $u$, the squared Yukawa coupling $ h^2$ and the anomalous dimensions $\eta_\phi,\eta_\psi$. To do so, we substitute our ansatz for $\Gamma_k$, Eq.~\eqref{eq:Gamma}, into the Wetterich equation, Eq.~\eqref{eqn:Wetterich}, and project it onto the different couplings. This leads to a set of coupled differential equations of the form
\begin{align}
\partial_t h^2&=\beta_{h^2}( h^2, u^{(n,m)}, \eta_\phi,\eta_\psi)\,, \\
\partial_t u&=\beta_{u}( h^2, u^{(n,m)}, \eta_\phi,\eta_\psi)\,,
\end{align}
together with two algebraic equations for $\eta_\phi$ and $\eta_\psi$. Here, $u^{(n,m)}$ denotes the derivative of the effective potential with respect to the invariants, i.e.
\begin{align}
u^{(n,m)}=\frac{\partial^{n+m}}{\partial \rho^n \partial \tau^m}u\,.
\end{align}

For the definition of the scale-dependent boson couplings from the effective potential, we expand $u$ in powers of $\rho$ and $ \tau$.
Generally, to properly access the fixed-point properties, we expand the effective fixed-point potential about its scale-dependent minimum.
In the considered model, it turns out that the minimum of the fixed-point potential typically lies at the origin $(\rho,\tau)=(0,0)$ except for very small $N_f\lesssim 1/2$.
We therefore choose an expansion corresponding to a vanishing vacuum expectation value of the bosonic field, which we refer to as the expansion in the symmetric (SYM) regime,
\begin{align}\label{eq:ueffexp}
	u(\rho,\tau)=&\sum_{i,j}\frac{1}{i!j!}\lambda_{i,j}\rho^i\tau^j\,.
\end{align}
This expansion includes the (dimensionless versions of the) previous couplings
\begin{align}
m^2=\lambda_{1,0},\quad g=\lambda_{0,1},\quad 8\lambda=\lambda_{2,0}\,.
\end{align}
The $\beta$-functions for the expansion parameters $\lambda_{i,j}$ in the SYM regime can then be obtained by appropriate projection prescriptions for the flow of the dimensionless effective potential $\partial_t u$. That is we obtain the flow of the bosonic couplings in the SYM regime from
\begin{align}
	\partial_t \lambda_{ij}=\left(\frac{\partial^{i+j}}{\partial \rho^i\tau^j}\partial_t u(\rho,\tau)\right)\Big|_{\rho=\tau=0}\,.
\end{align}
Accordingly, we set $ \rho=\tau=0$ in the expressions for $\partial_t h^2$, $\eta_\phi$ and $\eta_\psi$.
For the explicit fixed-point solutions, we expand the effective potential up to a finite order in the boson fields. That is,  it includes all symmetry-allowed couplings up to order $\phi^N$ where, in practice, $N\in \{4,6,8,12\}$. We can then also study the dependence of our results on $N$ to establish a convergence within the potential expansion. We refer to the approximation including the effective potential up to order $N$, together with the running wave function renormalizations and Yukawa coupling as LPA$N^\prime$ in the following. We explain further details of the derivation for the flow equations including projection prescriptions and regulator choices in App.~\ref{app:flow}. We also present the explicit and fully analytical expressions for the $\beta$~functions for an expansion with general vacuum expectation value of the field in App.~\ref{app:flow}.

\subsection{FRG flow equations in the symmetric regime}\label{sec:frgsub}

For vanishing expectation value the flow of the Yukawa coupling is completely given by its dimensional running
\begin{align}
\partial_t  h^2=&(D-4+\eta_\phi+2\eta_\psi) h^2\,,
\end{align}
because the loops with $\phi_1$ and $\phi_2$ propagator exactly cancel each other.

As example of the flow of the boson couplings, we list the flow equations of the lowest order couplings of  Eq.~(\ref{eq:ueffexp})
\begin{align}
\partial_t m^2&=(\eta_\phi-2)m^2+\frac{8v_D}{D}\left(1-\frac{\eta_\psi}{1+D}\right)N_f d_\gamma  h^2\\
&+\frac{32v_D}{D}\left(1-\frac{\eta_\phi}{2+D}\right)\left( \frac{9g^2}{(1+m^2)^3} - \frac{4\lambda}{(1+m^2)^2}\right)\,,\nonumber\\
\partial_t g&=\frac{1}{2}(D-6+3\eta_\phi)g\\
&+\frac{32v_D}{D}\left(1-\frac{\eta_\phi}{2+D}\right)\left(\frac{12\lambda g}{(1+m^2)^3}-\frac{\lambda_{1,1}}{(1+m^2)^2}\right)\,,\nonumber\\
\partial_t\lambda&=(D-4+2\eta_\phi)\lambda-\frac{4v_D}{D}\left(1-\frac{\eta_\psi}{1+D}\right)N_f d_\gamma h^4 \nonumber\\
&+\frac{16v_D}{D}\left(1-\frac{\eta_\phi}{2+D}\right)\Bigg(\frac{162g^4}{(1+m^2)^5}-\frac{216g^2\lambda}{(1+m^2)^4}\notag\\
&+\frac{40\lambda^2}{(1+m^2)^3}+\frac{18g\lambda_{11}}{(1+m^2)^3}-\frac{9}{8}\frac{\lambda_{02}}{(1+m^2)^2}\notag\\
&-\frac{3}{16}\frac{\lambda_{30}}{(1+m^2)^2}\Bigg)\,,\\
\partial_t\lambda_{11}&=\ldots\notag\,\\
\vdots\notag\,
\end{align}
Their graphical representation is given in Fig.~\ref{fig:feyn01}. At these expressions, one can recognize the FRG enhancement by the dressing of the propagators with the anomalous dimensions.
The anomalous dimensions are given by the solution of
\begin{align}
\eta_\psi&=\frac{16v_D}{D} h^2 \left(1-\frac{\eta_\phi}{D+1}\right)\frac{1}{(1+m^2)^2}\,,\\
\eta_\phi&=\frac{2v_D(4-3D+2\eta_\psi)}{D(2-D)} N_fd_\gamma  h^2+\frac{144v_D}{D}\frac{g^2}{(1+m^2)^4}\,.
\end{align}
%

\subsection{Emergent $U(1)$ symmetry}

We can now directly evaluate the renormalization group equations for arbitrary dimension $D$ and fermion flavor number $N_f$. This allows us to determine the fixed point structure and the critical behavior.
The fixed point corresponding to the potentially continuous Kekul\'e transition exhibits an enhanced $U(1)$ symmetry. This has also been observed in the perturbative RG calculations\cite{li2015,Scherer:2016zwz}.
Therefore, all couplings that reduce the symmetry to $\mathbb{Z}_3$ vanish at the fixed point.
More specifically, this allows us to reduce the set of couplings to be solved for in the fixed point equations.
We can directly use
\begin{align}
\lambda_{i,j}^\ast=0\quad \text{for}\quad j\neq0\,,
\end{align}
and solve for the NGFP solution in the $\lambda_{i,0}$ couplings.
For the stability matrix, however, all the couplings have to be taken into account as the RG scaling generally is different from zero, also for the vanishing couplings.
The diagonalization of the stability matrix is simplified by an effective decoupling of the $U(1)$ symmetric and the $U(1)$-symmetry-broken sector at the fixed point leading to a block diagonal structure.

\subsection{Connection to perturbative RG}

\begin{figure}[t]
\includegraphics[width=0.95\columnwidth]{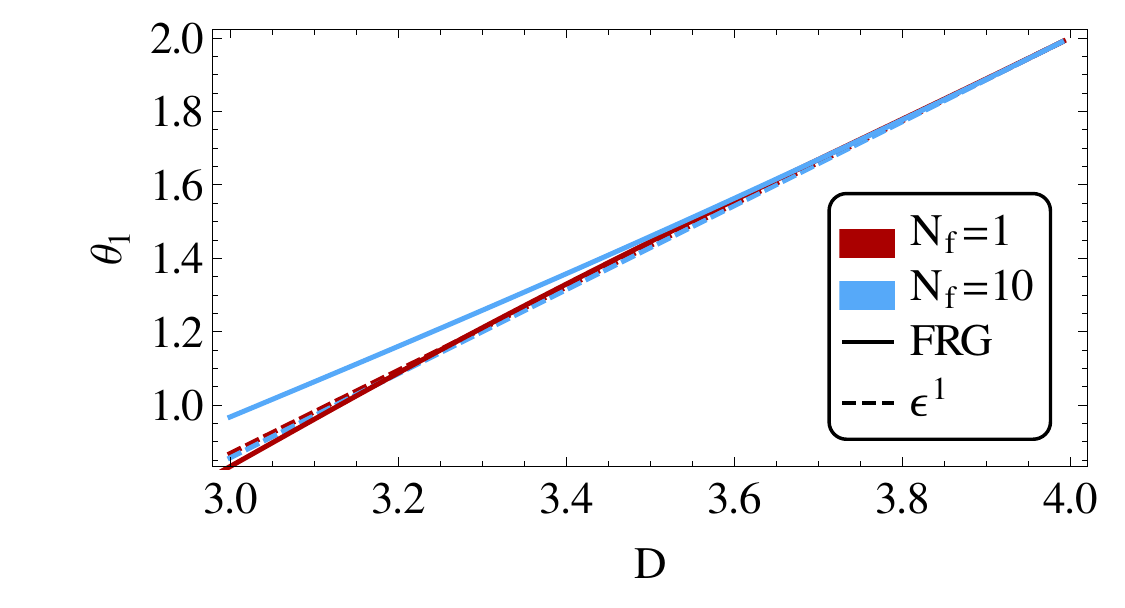}
\includegraphics[width=0.95\columnwidth]{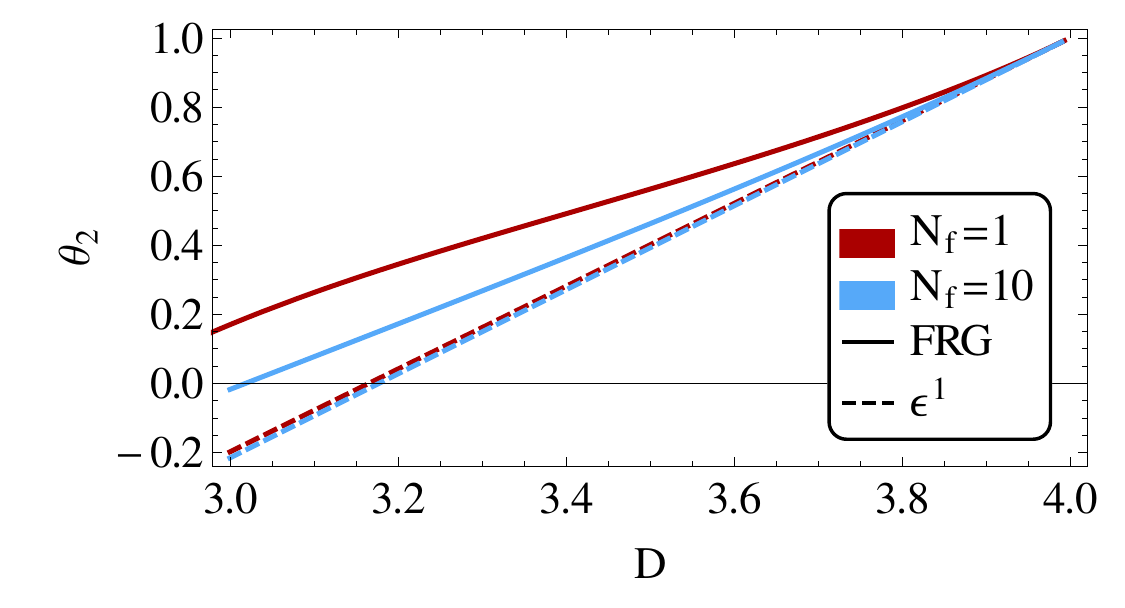}
\caption{The two largest critical exponents as function of the space-time dimension for fixed $N_f=1$ and $N_f=10$. Our FRG result is compared to an expansion around the upper critical dimension $D=4-\epsilon$ to first order in $\epsilon$. In contrast to this perturbative calculation, we find that the decisive second largest exponent remains positive in three space-time dimensions for $N_f=1$.}
\label{fig:D}
\end{figure}

In order to establish a connection to the perturbative RG approach, we compare our results for the universal critical exponents to a perturbative expansion in $4-\epsilon$ dimensions to first order in $\epsilon$, cf. Ref.~\onlinecite{Scherer:2016zwz}.
Since this expansion becomes exact close to the upper critical dimension, our FRG results should reduce to this limit\cite{janssen2014,Classen:2015mar,eichhorn2016}.
To confirm this, we have calculated the universal critical exponents for dimensions $D\in [3,4]$.
In Fig.~\ref{fig:D}, we present the cases $N_f=1$ and $N_f=10$, showing that the FRG results indeed approach the values of the $\epsilon$-expansion in the perturbative regime.
However, for dimensions close to three, which is relevant for two-dimensional Dirac materials, deviations increase.
In fact, the second critical exponent remains positive for $N_f=1$ in contrast to the $\epsilon$-expansion result.
For $N_f=10$, $\theta_2$ changes sign, but not before $D\approx3.01$. In this case, the $U(1)$ fixed point becomes stable in $D=3$ and renders the putative discontinuous Kekul\'e transition second order.
However in comparison to first order in $\epsilon$, the tendency to a fluctuation induced second order transition is still reduced.
This general trend can be attributed to the threshold effects from the FRG approach, i.e. the mass contributions in the propagators.
Thus the impact of fluctuations is reduced by the diminution of loop contributions.
In contrast, these effects are not included in a perturbative expansion to one-loop order, so that fluctuations there tend to be overestimated.

\subsection{Fermion-induced criticality}

\subsubsection{Fixed point stability for varying numbers of fermions}
\begin{figure}[t!]
\includegraphics[width=0.95\columnwidth]{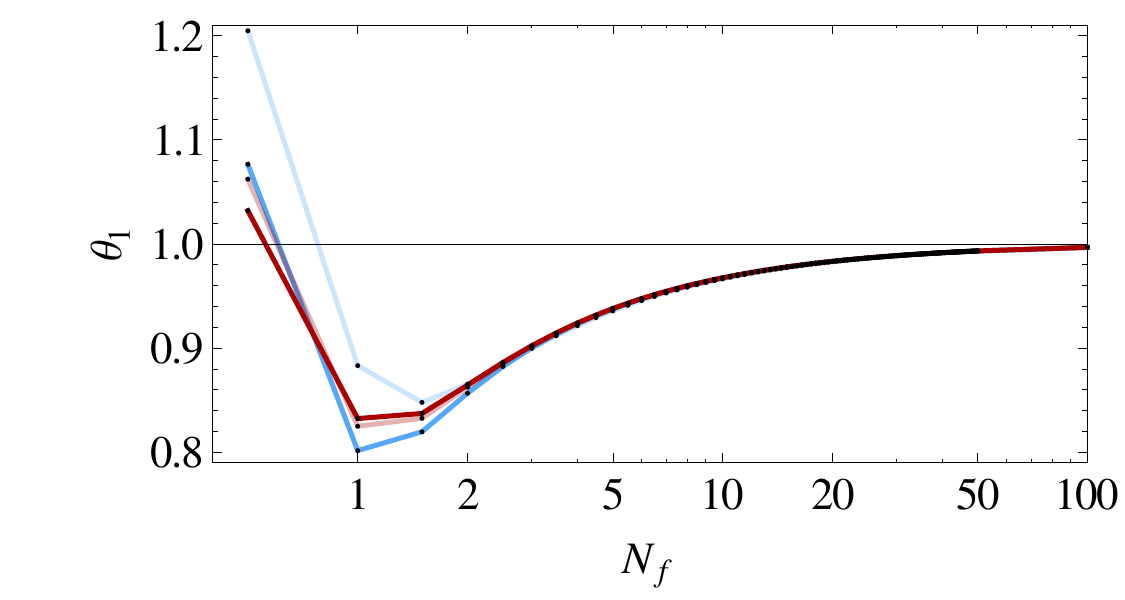}
\includegraphics[width=0.95\columnwidth]{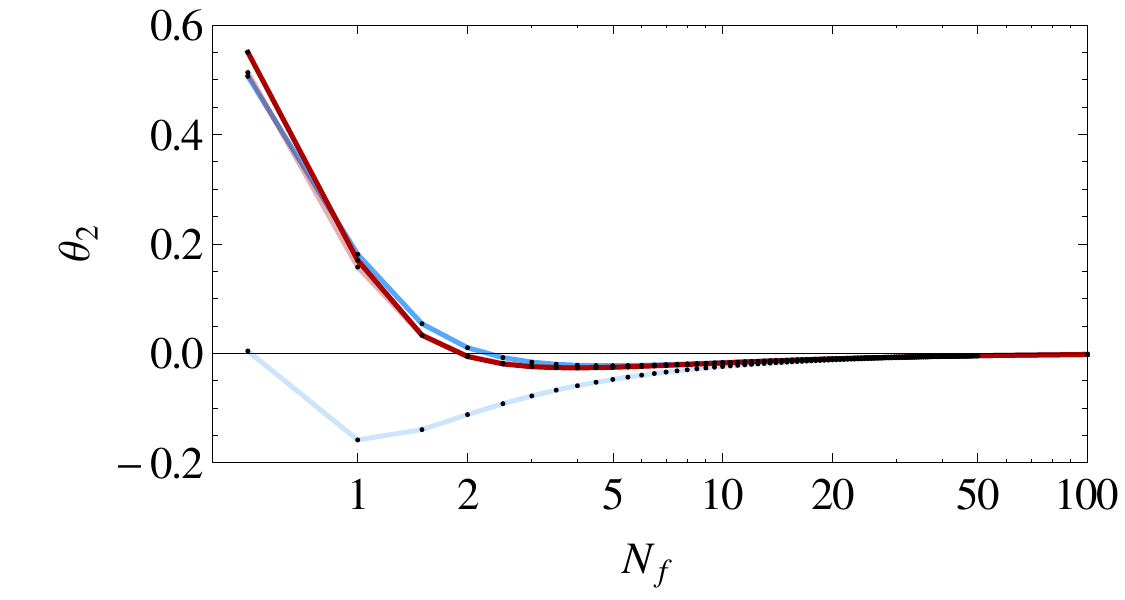}
\caption{Largest two critical exponents in $D=2+1$ for different $N_f$. The first critical exponent determines the correlation length exponent $\theta=1/\nu$. The second decides over the order of the phase transition. If $\theta_2<0$ the transition is continuous. The different lines mark different orders of the expansion: light blue, blue, red, light red corresponds to LPA4${}^\prime$, LPA6${}^\prime$, LPA8${}^\prime$ and LPA12${}^\prime$, respectively. LPA$N^\prime$ denotes that the order parameter potential is expanded up to the $\phi^N$ coupling. For large $N_f$, the critical exponents approach the values $\theta_i=D-(i+1)$, see text above Eq.~\eqref{eq:explargeNf}. \label{fig:Nf}
}
\end{figure}

To determine if the nature of the Kekul\'e phase transition can be changed from first to second order in the case of two-dimensional Dirac materials as graphene, we calculate the critical exponents characterizing the $U(1)$-symmetric fixed point in $D=2+1$ for different numbers of pairs of Dirac points $N_f$.
As we explained in Sec.~\ref{sec:frg}, we determine the second order regime by analyzing the sign of the second largest critical exponent, which decides over the stability of a fixed point. If it becomes negative, the phase transition becomes second order and universal critical behavior emerges.
The largest critical exponent is related to the correlation length exponent $\theta_1=\nu^{-1}$.
It is always positive.
With respect to the second largest exponent, we find that it changes sign at a critical $N_{f,c}$. Below $N_{f,c}$, it is positive in our calculations so that a first order transition takes place. Above the critical number of fermions, $\theta_2$ is always negative. In this regime, fluctuations are strong enough to change the sign of $\theta_2$ and make the $U(1)$ fixed point stable. All further critical exponents of the $U(1)$-symmetric fixed point are negative in our calculation.
Our best estimate for the critical fermion flavor number
is $N_{f,c}\approx1.9$. This lies in the range of upper and lower bounds obtained by QMC calculations\cite{li2015}, $N_{f,c}<2$, and an emergent SUSY theory\cite{2016arXiv161007603J,Zerf2016}, $N_{f,c}>1/2$.
Further, we show the largest critical exponents $\theta_1$ and $\theta_2$  as function of the number of fermions in Fig.~\ref{fig:Nf}.

To investigate the effect of higher order couplings, we have determined the critical behavior for different expansion orders of the order parameter potential. More specifically we have taken into account all symmetry-allowed couplings up to $\phi^{N}$ for $N \in \{4,6,8,12\}$. Interestingly, we find that for small $N_f$ a $\phi^4$ expansion as usually employed in perturbative calculations is not sufficient. It differs substantially from higher-order expansions as can be cleary seen in Fig.~\ref{fig:Nf}. In contrast, we observe a convergence of our results for orders higher than $\phi^6$. These findings confirm our expectations that in $D=2+1$ dimensions and for not too large fermion flavor numbers, the contribution of higher order couplings is not negligible, because couplings up to $\phi^6$ are relevant at the microscopic level. Although at the interacting fixed point higher order eigenperturbations become irrelevant, their scaling still contributes to the second critical exponent through non-diagonal entries in the stability matrix.
Finally, for increasing $N_f$, boson loops become sub-leading compared to fermion contributions so that the difference between the expansion orders of the boson potential stops playing a role.
%

\subsubsection{Large $N_f$}

For large $N_f$, we can solve the fixed point equations analytically. This allows us to obtain exact results for the critical exponents. In the symmetric regime for large $N_f$, the FRG equations simplify to
\begin{align}
	\partial_t u =& -D u +(D-2+\eta_\phi)( \rho u^{(1,0)}+\frac{3}{2}\tau u^{(0,1)})\nonumber \\ &-  \frac{4v_D d_\gamma}{D} \frac{1}{1+2 h^2\rho} + \mathcal{O}\left(\frac{1}{N_f}\right)\label{eq:ulargeNf}\,,\\
	\partial_t  h^2&=(D-4+\eta_\phi) h^2 + \mathcal{O}\left(\frac{1}{N_f}\right)\,,\\
	\eta_\phi=&\frac{8v_Dd_\gamma(4-3D)}{D(8-4D)}  h^2  + \mathcal{O}\left(\frac{1}{N_f}\right)\,,\\
\eta_\psi=& \mathcal{O}\left(\frac{1}{N_f}\right),
\end{align}
where we have rescaled $u\rightarrow u/N_f$, $Z_\phi\rightarrow Z_\phi/N_f$.
The fixed point that we are interested in has nonzero $ h^2$. Note, that the solution with $ h^2=0$ is unstable, because it has an additional relevant eigenperturbation in the direction of $ h^2$. Consequently, from $\partial_t  h^2=0$ follows that $\eta_{\phi}=4-D$.
Furthermore, $\partial_t  h^2$ only depends on $ h^2$ and
\begin{align}
\partial_t u^{(n,m)}
&=\left(-D+n+\frac{3}{2}m\right)u^{(n,m)} \notag \\
&\quad\quad- \frac{4v_D d_\gamma}{D}\left. \frac{\partial^{n+m}}{\partial \rho^n\partial \tau^m} \frac{1}{1+2 h^2 \rho}\right|_{ \rho= \tau=0}
\end{align}
only depends on $u^{(n,m)}$ and $ h^2$.
Therefore, the stability matrix becomes triangular and we can read off its eigenvalues from the diagonal entries.
That means the critical exponents are given by the scaling dimensions of the couplings at the interacting fixed point.
It follows for the boson couplings $\theta_i=[u^{(n,m)}]=D-(2n+3m)[\phi]=D-(2n+3m)\frac{1}{2}(D-2+\eta_\phi)$  and we obtain
\begin{align}
	\theta_i=D-2n-3m\,,
\end{align}
with $\eta_\phi=4-D$.
We show how the largest critical exponents approach these values in Fig.~\ref{fig:Nf}. In particular, we find for the cubic coupling
\begin{align}\label{eq:explargeNf}
\theta_2=[\lambda_{0,1}]=D-3.
\end{align}
Thus for large $N_f$, the cubic coupling is exactly marginal at the interacting fixed point leading to a line of fixed points for arbitrary $\lambda_{0,1}$.
We can also see this at the exact solution of Eq.~(\ref{eq:ulargeNf})
\begin{align}
u^*(\rho,\tau)=&-\frac{4 d_\gamma v_D}{D^2} {_2F_1}\left(1,-\frac{D}{2},1-\frac{D}{2},-2h^{2}_\ast\rho\right) \notag\\
&+\rho^{D/2} f\left(\frac{\tau}{\rho^{3/2}}\right)\,.
\end{align}
with the hypergeometric function ${_2F_1}$ and $h^{2}_\ast=D(8-6D+D^2)/(2v_Dd_\gamma(4-3D))$. The arbitrary function $f(\tau/\rho^{3/2})$ must be determined by boundary conditions. But we see that the cubic term $\lambda_{0,1}\tau=\lambda_{0,1}\rho^{3/2}\frac{\tau}{\rho^{3/2}}$  is of the form $\rho^{D/2} f(\tau/\rho^{3/2})$ in $D=3$. That is it satisfies the fixed point equation for every $\lambda_{0,1}$.
As a consequence of Eq.~\eqref{eq:explargeNf}, a second order phase transition can only occur for $D\leq3$ for large $N_f$.

\begin{figure}[t!]
\includegraphics[width=\columnwidth]{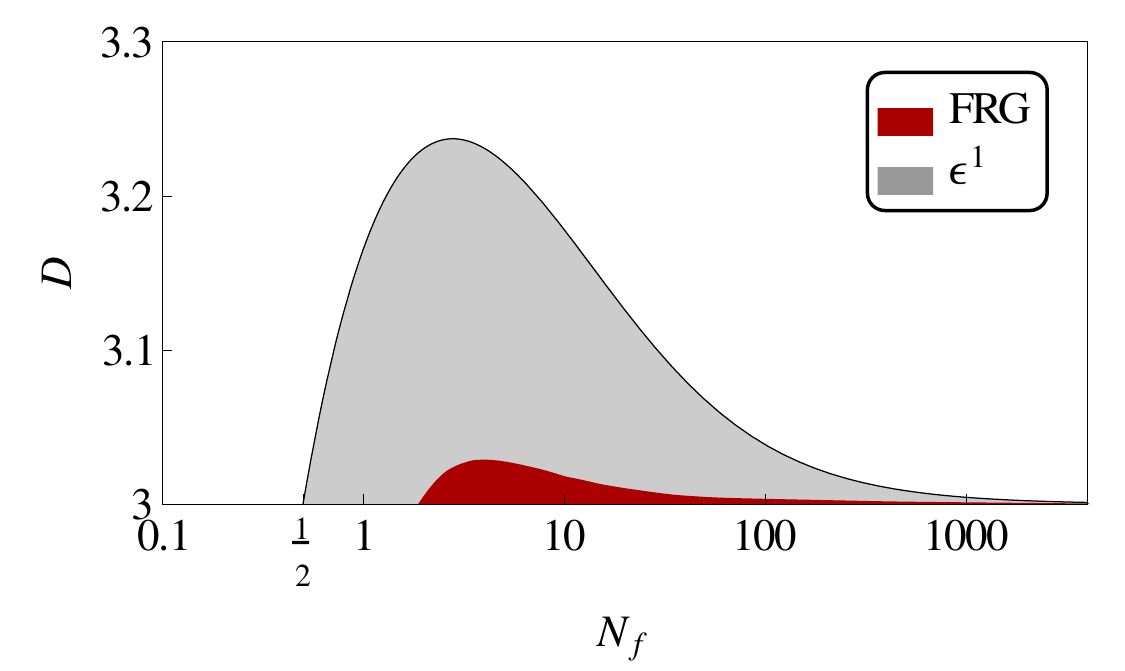}
\caption{Regime where the phase transition becomes of second order, for varying space-time dimension $D$ and fermion flavor number $N_f$. The red regime denotes the FRG result from this work. It is compared to the result obtained in the vicinity of the upper critical dimension $D=4-\epsilon$ \cite{Scherer:2016zwz}.}
\label{fig:DNf}
\end{figure}

\subsubsection{Nature of the phase transition for general $D$ and $N_f$}

We have also performed the stability analysis of the fixed point which
describes the phase transition
to the $\mathbb{Z}_3$-symmetric Kekul\'e state for
varying spacetime dimensions. 
This allows to estimate how close the theory in 2+1 spacetime dimensions is to the other regime,
i.e. if the transition is weakly first or second order, respectively.
Our result is shown in Fig.~\ref{fig:DNf}. The second order regime appears close to three space-time dimensions for $N_f\gtrsim1.9$. Compared to the first order $\epsilon$-expansion \cite{Scherer:2016zwz}, it is severely reduced. As we explained above, the reason is that higher order couplings play a role and that threshold effects coming from the full propagators in the fRG equations reduce fluctuation effects. Nevertheless, systems with $N_f=2$ and $D=2+1$ still lie in the range with a stable fixed point. That means that, e.g., spin-1/2 fermions on the two-dimensional honeycomb lattice show a second order transition with critical scaling. The scaling is determined by the correlation length exponent $\theta_1$, the exponent of the cubic anisotropy $\theta_2$ and the anomalous dimensions $\eta_i$ see tables~\ref{tab:eigen} and~\ref{tab:critexp1}. However, as a result of being near to the first order transition, we can expect substantial corrections to this scaling in the whole second order regime as explained in the next section.

\subsection{Critical behavior in 2D Dirac materials}

\subsubsection{Corrections to scaling}


\begin{table}[t!]
\caption{\label{tab:eigen} Numerical values for the largest three critical exponents for different $N_f$ in $D=2+1$ in LPA$12^\prime$. The inverse correlation length exponent is given by $\theta_1=\nu^{-1}$.The critical exponent deciding over stability $\theta_2$ is printed in boldface and also determines the corrections to scaling in the quantum critical regime.}
\begin{tabular*}{\linewidth}{@{\extracolsep{\fill} } l l l l}
\hline\hline
  $N_f$ & $\theta_1$ & $\theta_2$ & $\theta_3$ \\ \hline
1/2 & 1.0620 & {\bf +0.5130} & -0.9021 \\
1 & 0.8248 & {\bf +0.1574} & -0.8044 \\
2 & 0.8623 & {\bf -0.00497} & -0.8779 \\
3 & 0.9017 & {\bf -0.02398} & -0.9203 \\
4 & 0.9237 & {\bf -0.02646} & -0.9413 \\
5 & 0.9377 & {\bf -0.02560} & -0.9536 \\
10 & 0.9672 & {\bf -0.01793} & -0.9775 \\
20 & 0.9831 & {\bf -0.01054} & -0.9889 \\
50 & 0.9931 & {\bf -0.00465} & -0.9956 \\
$\infty$ & 1 & {\bf 0} & -1\\
 \hline\hline
\end{tabular*}
\end{table}

We list our best numerical values for the critical exponents, i.e. the eigenvalues of the stability matrix, as obtained within LPA$12^\prime$ in Tab.~\ref{tab:eigen}.
In the physical case of 2+1 dimensions, we see from Tab.~\ref{tab:eigen} and Fig.~\ref{fig:Nf} that the magnitude of $\theta_2$ is generally small, i.e. $|\theta_2|\ll 1$. This implies that there is an RG direction which flows very slowly, i.e. almost logarithmically, towards its quantum critical point.
Therefore, we expect that the corrections to scaling should be easily visible in the vicinity of the quantum critical point.

Generally, the leading correction to scaling at a continuous phase transition is induced by the least irrelevant RG direction at the critical point. Here, this RG direction is related to the cubic coupling $g$ and indeed represented by the critical exponent $\theta_2$ which gives rise to a power-law correction to the scaling.
For example, the scaling of the correlation length should follow a behavior of the form\cite{wegner1972b,herbutbook,assaad2013}
\begin{align}\label{eq:corr}
	\xi \sim A|\Delta|^{-\nu}\Big(1+B|\Delta|^{-\theta_2}+\ldots\Big)\,,
\end{align}
where $A,B$ are non-universal amplitudes, and in particular $B\propto g$. The ellipsis indicate further corrections due to the stronger irrelevant directions,  which vanish very quickly; see the values of $\theta_3$ in Tab.~\ref{tab:eigen}. $\Delta$ marks the distance to the critical point. In general, it represents the difference of the tuning parameter and its critical value, e.g. a coupling strength in a lattice description or the reduced temperature at a thermal phase transition.

Eq.~(\ref{eq:corr}) shows that for small negative $\theta_2$ the term proportional to the cubic coupling $\propto g$ will vanish only very slowly.
This behavior should be clearly detectable in a Quantum Monte Carlo simulation.
The corrections to scaling will become particularly important in the limits $N_f \gtrsim N_{f,c}$ and $N_f \to \infty$ as in both cases $\theta_2 \to 0^-$.

\subsubsection{Emergent SUSY fixed point}

\begin{table}[t!]
\caption{\label{tab:critexp2} Correlation length exponent $\nu=\theta_1^{-1}$ and anomalous dimensions for the emergent SUSY scenario, i.e. $N_f=1/2$ in the $U(1)$ symmetric model. The epsilon expansion results ($\epsilon^3$) to three-loop order are taken from Ref.~\onlinecite{Zerf2016} with direct substitution $\epsilon=1$. The conformal bootstrap (cBS) results have been calculated in Ref.~\onlinecite{Bobev:2015vsa}. The FRG results are obtained within LPA12${}^\prime$.}
\begin{tabular*}{\linewidth}{@{\extracolsep{\fill} } l l l l}
\hline\hline
$N_f=1/2$  & $\nu$ & $\eta_\phi$ & $\eta_\psi$ \\ \hline
 $\epsilon^3$  & 0.985 & 1/3 & 1/3\\
 cBS & 0.917 & 1/3 & 1/3 \\
 FRG & 0.941 & 0.354 & 0.323\\
 \hline\hline
\end{tabular*}
\end{table}

To obtain an estimate of the accuracy of our results, we compare them to several other methods.
In particular, there exist exact results in the limit where we set $N_f=1/2$ and restrict the symmetry to $U(1)$, which is the emergent symmetry at the considered fixed point.
In that case, the model under investigation exhibits the same universal critical behavior as the  Lagrangian describing the semimetal-superconductor quantum phase transition of a two-component Dirac fermion on the 2D surface of a 3D topological insulator\cite{hasan2010,qi2011} reading \cite{Roy:2012wz,Zerf2016}
\begin{align}\label{eq:susy}
	\mathcal{L}=&i\bar\psi\slashed{\partial}\psi+|\partial_\mu \phi|^2+m^2|\phi|^2+\lambda|\phi|^4\\
	&+h(\phi^\ast\psi^T i \sigma_2 \psi+\mathrm{h.c.})\nonumber\,,
\end{align}
where the gamma matrices are defined using the Pauli matrices, i.e. $\gamma_0=\sigma_3, \gamma_1=\sigma_1, \gamma_2=\sigma_2$, and $\bar\psi=-i\psi^\dagger\gamma_0$.
For the model in Eq.~\eqref{eq:susy}, an emergent SUSY scenario described by Wess-Zumino theory\cite{wess1974} at the critical point was discussed\cite{lee2007}. Quantitative estimates for the critical exponents have been suggested by means of the conformal bootstrap\cite{Bobev:2015vsa} and within the $4-\epsilon$ expansion up to three loop order\cite{Zerf2016}. We present the corresponding comparison of the correlation length exponent and the anomalous dimensions in Tab.~\ref{tab:critexp2}. Our results for the correlation length exponent lies within 5\% of the conformal bootstrap and the three-loop calculations. The anomalous dimensions agree within 6\% to the exact result of $\eta_\psi=\eta_\phi=1/3$.
Let us also note a technical aspect here. Small values of $N_f$ reduce the impact of fermion fluctuations, which will shift the fixed point value of the $\abs{\phi}^2$ coupling to negative values. This occurs around $N_f\approx 0.6$ in our calculations. In this case one can adapt the expansion of the effective potential to the new minimum appearing away from $\rho=\tau=0$ to obtain better estimates of critical exponents. We refrain from doing this here, because it only affects the $N_f=1/2$ case, where we nevertheless find reasonable critical exponents as demonstrated above.
%

\subsubsection{Comparison with MQMC}

For larger fermion flavor numbers, we can compare our results with quantum Monte Carlo calculations\cite{li2015} in $D=2+1$. We list the correlation length exponent and the boson anomalous dimension in Tab.~\ref{tab:critexp1}. In addition, we quote the results of a first order expansion around the upper critical dimension. Our anomalous dimensions lie within 19\% for $N_f=2$ to 8\% for $N_f=6$ of the MC results. In this respect, it is  known, however, that the FRG at this truncation order does not lead to very accurate boson anomalous dimensions. In contrast, we find that the correlation length exponents of our FRG calculation and the QMC result agree within 11\% for $N_f=2$, which is improved to 2\% agreement for $N_f=6$.
Let us note here, that our prediction of substantial corrections to scaling have not yet been considered in the QMC simulation so that the agreement between the critical exponents may even improve.

In conclusion of the comparison, we believe that we can give good estimates for the critical behavior for a broad range of $D$ and $N_f$. We reproduce the perturbative RG results close to $D=4$. Furthermore, our critical exponents are comparable to available quantitative approaches in $D=3$. Thus we are confident that our analysis provides valid results for the whole range of $D$ and $N_f$, which has not been available to this extent before.


\begin{table}[t]
\caption{\label{tab:critexp1} Critical exponents of the Kekul\'e transition for different $N_f$ in $D=2+1$. The QMC values and 1-loop RG results are taken from Ref.~\onlinecite{li2015}, the FRG results are calculated in LPA12${}^\prime$. We added our calculation of $\eta_\psi$ for completion.}
\begin{tabular*}{\linewidth}{@{\extracolsep{\fill} } l| l l l | l l l | l}
\hline\hline
$N_f$  & & $\nu$ & & & $\eta_\phi$ & & $\eta_\psi$   \\ \hline
 & QMC & $\epsilon^1$ & FRG  & QMC & $\epsilon^1$ & FRG &  FRG\\ \hline
$2$ & 1.04 &  1.25 &  1.16  & 0.71(3) & 0.67 & 0.88  &0.062\\
 $3$  & 1.05 &  1.26 &  1.11 & 0.77(2) & 0.75 & 0.92   & 0.038 \\
  $4$ & 1.12 & 1.25 & 1.08  & 0.80(4) & 0.80 & 0.95  &0.027\\
 $5$  & 1.08 &  1.23 &  1.07  & 0.85(4) & 0.83 & 0.96 &0.021 \\
  $6$ & 1.07 & 1.22 & 1.06  & 0.89(4) & 0.86 & 0.97 & 0.017\\
 \hline\hline
\end{tabular*}
\end{table}

\section{Conclusions}\label{sec:conc}

In this work, we have studied the quantum phase transition to the Kekul\'e phase in two-dimensional Dirac materials.
The condensation of the Kekul\'e order parameter reduces the ``chiral'' $U(1)$ symmetry of Dirac materials Eq.~(\ref{eq:chiralU1}) to $\mathbb{Z}_3$.
For electrons on the honeycomb lattice, it corresponds to a structural transition, where the translational invariance is broken because a valence bond solid is formed. The Kekul\'e VBS has already been observed in molecular graphene\cite{gomes2012} and graphene on a Copper substrate\cite{Gutierrez2016}. 
With respect to graphene on a substrate, it is interesting to consider the effect of doping. A chemical potential introduces a new scale to the system and leads to a finite density of states, in favor of chiral symmetry breaking. It also preserves the chiral symmetry and the nesting close to the two inequivalent Dirac cones. But the nesting vector is not commensurate to the lattice anymore, which is disadvantageous for the Kekulé order. Furthermore, a finite chemical potential destroys Lorentz symmetry. However, universal critical behavior is determined by the scaling of the Fermi surface, not by its absolute magnitude, and the behavior at zero doping is continuously connected to the one at finite doping. Thus we expect that the infrared behavior of the weakly doped case is still governed by the same fixed point as the neutral system as long as long-range order can still develop.

Within a conventional Landau-Ginzburg description in terms of an order parameter field with small fluctuations, the corresponding phase transition would be of first order. The reason is that cubic terms of the order parameter field are allowed by symmetry and consequently appear in the free energy. Interestingly, the order of the phase transition can be changed when fluctuations become strong. This happens for example in the closely related three-states Potts model in 1+1 dimensions\cite{baxter1973,RevModPhys.54.235}, where the effect of fluctuations is increased due to the reduced dimensionality. Here, we have considered another intriguing mechanism to induce a second order transition, which is specific to the quantum character of the phase transition. At zero temperature, the Dirac fermions, which are massless in the symmetric phase, provide an additional critical mode at the phase transition. That means that fluctuations are increased due to the presence of further degrees of freedom, besides the order parameter field. We have investigated if this can indeed lead to a change of the nature of the Kekul\'e transition in Dirac materials.

We have modeled the system in terms of a Gross-Neveu-Yukawa theory with a generalized number of Dirac fermions $N_f$. Thereby, for example, $N_f=2$ describes graphene, and $N_f=1$ corresponds to spinless fermions on the honeycomb lattice. Further, we have discussed the emergent SUSY scenario for the case of $N_f=1/2$. With the help of the non-perturbative functional renormalization group, we have derived flow equations for the parameters of the model and investigated the fixed point structure. The fixed point corresponding to the Kekul\'e transition shows an emergent $U(1)$ symmetry and its stability depends on the number of Dirac fermions. Stability means the ability to tune to the critical point, which in turn decides if the transition is of first or second order. We found that the nature of the transition changes when $N_f\approx1.9$. That means that the physically interesting case of graphene, or graphene-like systems, displays a second order phase transition. Hence, if the Kekul\'e transition becomes experimentally accessible, critical behavior should be observable.

The critical behavior at a second order phase transition is revealed by the scaling of correlation functions, characterized by critical exponents. We have calculated these critical exponents for the second order regime. Due to the proximity to the first order transition, the second largest critical exponent is, albeit irrelevant, close to zero. For large $N_f$, it further tends to exactly zero, i.e.~the associated coupling, which corresponds to the cubic term of the order parameter field, is exactly marginal. We expect that this affects the scaling laws close to the phase transition, because they are modified by the scaling of the (almost) marginal coupling. The effect should be clearly visible in QMC simulations.

We also showed that for an accurate determination of the critical $N_f$ and the critical exponents, it is crucial to account for higher order couplings in the continuum description. In perturbative calculations, couplings of higher order than $\phi^4$ are usually omitted. But in 2+1 dimensions, couplings up to $\phi^6$ are relevant at the bare level. Although they turn irrelevant at the interacting fixed point, which describes the Kekul\'e transition, their scaling affects the critical exponents.

A further aspect of the critical behavior of the Kekul\'e transition is that through the breaking of the continuous $U(1)$ symmetry to $\mathbb{Z}_3$ a second scale is introduced. The reason is that the Goldstone mode of the $U(1)$ symmetry acquires a dynamical mass in the symmetry-broken phase, which is different from the longitudinal mass. Hence, the scaling of correlations is different on the two sides of the phase transition \cite{Leonard:2015wyg,2016arXiv161007603J}. Interestingly, the FRG allows to study the dynamical generation of a mass gap in the spontaneously broken regime. We leave it for future work to demonstrate the different scaling in the Kekul\'e and the Dirac semimetal phase.

\begin{acknowledgments}
The authors are grateful to E.~Torres and Andr\'es Goens for discussions. IFH is supported by the NSERC of Canada.
This work has been supported by the Deutsche Forschungsgemeinschaft (DFG) through the Collaborative Research Center SFB 1238, TP C04. LC is supported by the U.S. Department of Energy (DOE), Division of Condensed Matter Physics and Materials Science, under Contract No.  DE-AC02-98CH10886.
\end{acknowledgments}

\begin{widetext}

\appendix

\section{FRG flow equations}\label{app:flow}

In this appendix, we present the derivation and the full expressions for the FRG flow equations. That means in our truncation for the effective action Eq.~(\ref{eq:Gamma}), we obtain flow equations for $Z_{\psi,k}, Z_{\phi, k}, \bar h_k$ and $U_k$. We express them in terms of the corresponding dimensionless quantities defined in Eqs.~(\ref{eq:dimlessU})-(\ref{eq:etas}).
For convenience, here and in the following, we suppress the index $k$.

\subsection{Effective potential}

The flow of the effective potential is obtained by evaluating Eq.~\eqref{eqn:Wetterich} for constant bosonic field $\phi$ and vanishing fermion field $\psi$. This allows us to present a closed form for the flow of the full effective potential within the given truncation. It reads
\begin{align}\label{eq:ueff}
	\partial_t u = -D u +(D-2+\eta_\phi)(\rho u^{(1,0)}+\frac{3}{2}\tau u^{(0,1)}) +2v_D l_0^{(B)}\left(m_L^2,\eta_\phi\right)+2v_D l_0^{(B)}\left(m_T^2,\eta_\phi\right)-2v_D N_f d_\gamma l_0^{(F)}(2h^2\rho,\eta_\psi)
\end{align}
with the volume element $v_D^{-1}=2^{D+1}\pi^{D/2}\Gamma[D/2]$. We have denoted the derivatives with respect to the invariants by $u^{(i,j)}=\frac{\partial^{i+j}}{\partial \rho^i\partial \tau^j} u$.
We have furhter introduced the longitudinal and transverse masses $m^2_{L/T}$ and the threshold functions $l_0^{(B)}$ and $I_0^{(F)}$. The threshold functions involve the loop integrations and the corresponding regulator dependence. We give the exact expressions in the App.~\ref{app:thresholds}.
The longitudinal and transversal mass $m_L^2$ and $m_T^2$ are obtained by diagonalizing $(\Gamma_k^{(2),B}+R_k^B)^{-1}$ and read $m_{L/T}^2=(u_{11}+u_{22})/2\pm\sqrt{4u_{11}^2+(u_{11}-u_{22})^2}/2$ as function of $\phi_1,\phi_2$ with $u^{(i,j)}=\frac{\partial^i}{\partial \rho^i}\frac{\partial^j}{\partial \tau^j} u$. When expressed in terms of $\rho,\tau$ the expressions become
\begin{align}
m_{L/T}^2&=+\rho  u^{(2,0)}(\rho ,\tau )+u^{(1,0)}(\rho ,\tau )+3 \tau  u^{(1,1)}(\rho ,\tau )+ 9 \rho ^2 u^{(0,2)}(\rho ,\tau )\nonumber \\
&\pm\Bigg[81 \rho ^4 u^{(0,2)}(\rho ,\tau )^2+\rho  \left(36 \rho ^2 u^{(1,1)}(\rho ,\tau )^2+6 \tau  u^{(2,0)}(\rho ,\tau ) u^{(1,1)}(\rho ,\tau
   )+\rho  u^{(2,0)}(\rho ,\tau )^2\right)\nonumber \\
   &\quad\quad+6 u^{(0,1)}(\rho ,\tau ) \left(12 \rho ^2 u^{(1,1)}(\rho ,\tau )+9 \rho  \tau  u^{(0,2)}(\rho ,\tau )+\tau  u^{(2,0)}(\rho ,\tau
   )\right)\nonumber \\
   &\quad\quad+9 u^{(0,2)}(\rho ,\tau ) \left(\left(\tau ^2-2 \rho ^3\right) u^{(2,0)}(\rho ,\tau )+6 \rho ^2 \tau  u^{(1,1)}(\rho ,\tau )\right)+36 \rho  u^{(0,1)}(\rho ,\tau
   )^2\Bigg]^{1/2}\,.
\end{align}

As we have explained in the main text, we expand $u$ in powers of $\rho$ and $\tau$ about its scale-dependent minimum. In our case the minimum of the fixed-point potential lies at the origin $(\rho,\tau)=(0,0)$ (except for very small $N_f\lesssim 1/2$).

\subsection{Yukawa coupling}

To extract the flow equation of the Yukawa coupling we decompose the two-point function into its fluctuation dependent and independent parts
$
\Gamma_{k,0}^{(2)}=\Gamma_k^{(2)}|_{\Delta\phi=\psi=0},
$ and
$
\Delta\Gamma_k^{(2)}=\Gamma_k^{(2)}-\Gamma_{k,0}^{(2)}\,
$
with $\phi=\phi_0+\Delta\phi$.
Then, we expand the Wetterich equation as
\begin{align}\label{eqn:logWetterich}
\partial_t\Gamma_k =& \frac{1}{2}\tilde\partial_t\text{STr}[\ln(\Gamma_k^{(2)}+R_k)]=\frac{1}{2}\tilde\partial_t\text{STr}[\ln(\Gamma_{k,0}^{(2)}+R_k)]  \nonumber+ \frac{1}{2}\tilde\partial_t\text{STr}\sum_{n=1}^\infty\frac{(-1)^{n+1}}{n}[(\Gamma_{k,0}^{(2)}+R_k)^{-1}\Delta\Gamma_k^{(2)}]^n\nonumber\,,
\end{align}
where we have defined the scale derivative $\tilde\partial_t$ which acts only on the $t$-dependence of the regulator (see also Eq.~(\ref{eq:scalederiv})).
The real and imaginary components of the bosonic field are also divided into their vacuum expectation value and a fluctuating part, $\phi_1=\bar\phi_1+\Delta\phi_1$ and $\phi_2=\bar\phi_2+\Delta\phi_2$.
To project the flow equation onto the Yukawa coupling, we then use
\begin{align}
\partial_t h = \frac{-i}{N_f d_\gamma}\text{Tr}\left[ \gamma_3 \frac{\delta}{\delta \Delta \phi_1(p')} \frac{\delta}{\delta \bar \psi(p)} \partial_t \Gamma_k \frac{\delta}{\delta \psi(q)}\right]\,,
\end{align}
evaluated at $\bar\psi=\psi=0$, $\Delta\phi_1=\Delta\phi_2=0$ and $p'=p=q=0$.
The flow of the squared Yukawa coupling can then be expressed in terms of field expectations values $\bar\phi_1, \bar\phi_2$. We rewrite them in terms of the invariants at the minimum of the effective potential $\rho_0=(\bar\phi_1^2+\bar\phi_2^2)/2$, $\tau_0=(\bar\phi_1^3-3\bar\phi_1\bar\phi_2^2)/\sqrt{2}$. Therefore we choose our coordinate system so that the minimum lies on the $\phi_1$ axis, i.e. $\bar\phi_2=0$. Then, if we denote the value of $\rho$ at the minimum as $\rho_{0}=\kappa$, we have the relation $\kappa=\bar\phi_1^2/2$ and $\tau_0=\sqrt{2\kappa^3}$. With this the $\beta$-function of the squared Yukawa coupling becomes
\begin{align}
\partial_t h^2=&(D-4+\eta_\phi+2\eta_\psi)h^2-8v_D h^4\left(l_{11}^{(FB)}(2h^2\kappa,m_{T,0}^2;\eta_\psi,\eta_\phi) + l_{11}^{(FB)}(2h^2\kappa m_{L,0}^2;\eta_\psi,\eta_\phi)\right) \nonumber \\[5pt]
&-8v_D\sqrt{2\kappa}h^4\omega_{111} l_{12}^{(FB)}(2h^2\kappa,m_{L,0}^2;\eta_\psi,\eta_\phi)
+8v_D \sqrt{2\kappa}h^4 \omega_{221} l_{12}^{(FB)}(2h^2\kappa,m_{T,0}^2;\eta_\psi,\eta_\phi)  \nonumber \\[5pt]
&+32 v_D\kappa h^6  l_{21}^{(FB)}(2h^2\kappa,m_{T,0}^2;\eta_\psi,\eta_\phi)-32v_D\kappa h^6 l_{21}^{(FB)}(2h^2\kappa,m_{L,0}^2;\eta_\psi,\eta_\phi) \,,
\end{align}
with $\omega_{ijk}=\frac{\partial^3}{\partial \phi_i \partial \phi_j \partial \phi_k} u\big|_{min}$ and $m_{L/T,0}^2=m_{L/T}^2\big|_{\text{min}}$ evaluated at the minimum of the potential.
They read
\begin{align}\label{eq:masses}
m_{L,0}&=u^{(1,0)}+2\kappa u^{(2,0)}+ 12\kappa^{3/2}u^{(1,1)} +6\kappa^{1/2}u^{(0,1)} + 18\kappa^2u^{(0,2)}\,,\\
m_{T,0}&=u^{(1,0)}-6\sqrt{\kappa}u^{(0,1)}\,,\\
\omega_{111}&=54 \sqrt{2} \kappa ^{5/2} u^{(1,2)}+54 \sqrt{2} \kappa ^{3/2}
   u^{(0,2)}+2 \sqrt{2} \kappa ^{3/2} u^{(3,0)}+54
   \sqrt{2} \kappa ^3 u^{(0,3)}+18 \sqrt{2} \kappa ^2 u^{(2,1)}+27 \sqrt{2} \kappa  u^{(1,1)}\nonumber\\
   &+3 \sqrt{2} \sqrt{\kappa }
   u^{(2,0)}+3 \sqrt{2} u^{(0,1)}\,,\\
\omega_{221}&=\sqrt{2} \left(-18 \kappa ^{3/2} u^{(0,2)}-3 \kappa  u^{(1,1)}+\sqrt{\kappa } u^{(2,0)}-3 u^{(0,1)}\right).
\end{align}
In our case these expressions simplify even further, because $\kappa=0$ in the SYM regime.
The threshold functions $l_{ij}^{(FB)}$ required for the Yukawa coupling are also displayed in App.~\ref{app:thresholds}.

\subsection{Anomalous dimensions}

To determine the expressions for the anomalous dimensions of the boson and fermion fields, we first need appropriate projection prescriptions for the wave function renormalizations.
To that end, we evaluate the Wetterich equation for momentum-dependent fields and choose the following prescriptions
\begin{align}
\partial_t Z_\phi&=\frac{\partial}{\partial p^2}\int_q \left .\frac{\delta}{\delta\phi_2(-p)} \frac{\delta}{\delta\phi_2(q)} \partial_t \Gamma_k \right|_{\substack{\bar \psi=\psi=0\\ \Delta\phi_2=0 \\ p=q=0}}\,, \\
\partial_t Z_\psi&=\frac{-i}{N_fd_\gamma D}\text{Tr}\left[\gamma_\mu\frac{\partial}{\partial p_\mu}\int_q \frac{\delta}{\delta\bar\psi(p)}  \partial_t \Gamma_k \frac{\delta}{\delta\psi(q)}\right]_{\substack{\Delta\phi_i=0 \\ p=q=0}}\nonumber\,,
\end{align}
with short-hand notation $\int_q=\int \frac{d^Dq}{(2\pi)^D}$.
The evaluation of these prescriptions yields a fermion anomalous dimension reading
\begin{align}
\eta_\psi=\frac{8v_D}{D}h^2 \left( m_{12}^{(FB)}(2h^2\kappa,m_{L,0}^2;\eta_\psi,\eta_\phi) + m_{12}^{(FB)}(2h^2\kappa,m_{T,0}^2;\eta_\psi,\eta_\phi) \right)\,.
\end{align}
Similarly, the boson anomalous dimension is given by:
\begin{align}
\eta_\phi&=\frac{4v_d}{d}\omega_{111}^2m_4^{(B)}(m_{L,0}^2,\eta_\phi) + \frac{4v_d}{d}\omega_{221}^2m_4^{(B)}(m_{T,0}^2,\eta_\phi) + \frac{8v_d}{d}N_f d_\gamma h^2 m_4^{(F)}(2h^2\kappa,\eta_\psi) - \frac{16v_d}{d}N_f d_\gamma \kappa h^4  m_2^{(F)}(2h^2\kappa,\eta_\psi)
\end{align}
and the threshold functions $m_{12}^{(FB)},m_{22}^{(B)},m_{4}^{(F)}$ and $m_2^{(F)}$ are again listed in App.~\ref{app:thresholds}.
This completes our list of FRG equations.

\subsection{Flow equations in the symmetric regime}

In the main text, we have stated the above expressions in the symmetric regime, where they simplify significantly. To obtain them, we have set $\kappa=0$ and $m_{L,0}^2=m_{T,0}^2=u^{(1,0)}=m^2$ and used the expressions for the threshold functions with the linear cutoff as given in the next section.

\section{Threshold functions} \label{app:thresholds}

In the above flow equations, we have abbreviated the result of the loop integrations in terms of threshold functions. These functions also contain the regulator dependence of the FRG equations. They are defined as
\begin{align}
l_0^{(B)}(\omega,\eta_\phi)&=\frac{1}{4v_d}\int \frac{d^dp}{(2\pi)^d} p^2 \frac{\partial_t r_\phi -\eta_\phi r_\phi}{p^2(1+r_\phi)+k^2\omega}\,,\\
l_0^{(F)}(\omega,\eta_\psi)&=\frac{1}{2v_d}\int \frac{d^dp}{(2\pi)^d} p^2 \frac{(1+r_\psi)(\partial_t r_\psi -\eta_\psi r_\psi)}{p^2(1+r_\psi)^2+k^2\omega}\,,\\
l_{nm}^{(FB)}(\omega_\psi,\omega_\phi;\eta_\psi,\eta_\phi)&=-\frac{1}{4v_d}k^{2(n+m)-d}\tilde\partial_t\int \frac{d^dp}{(2\pi)^d}\frac{1}{(p^2(1+r_\psi)^2+k^2\omega_\psi)^n(p^2(1+r_\phi)+k^2\omega_\phi)^m}\,,\\
m_4^{(B)}(\omega,\eta_\phi)&=-\frac{1}{4v_d}k^{6-d}\tilde\partial_t\int \frac{d^dp}{(2\pi)^d}p^2\left( \frac{\partial}{\partial p^2}\frac{1}{p^2(1+r_\phi)+k^2\omega} \right)^2\,,\\
m_{22}^{(B)}(\omega_1,\omega_2;\eta_\phi)&=-\frac{1}{4v_d}k^{6-d}\tilde\partial_t\int \frac{d^dp}{(2\pi)^d}p^2\left( \frac{\partial}{\partial p^2}\frac{1}{p^2(1+r_\phi)+k^2\omega_1} \right)\left( \frac{\partial}{\partial p^2}\frac{1}{p^2(1+r_\phi)+k^2\omega_2} \right)\,,\\
m_4^{(F)}(\omega,\eta_\psi)&=-\frac{1}{4v_d}k^{4-d}\tilde\partial_t\int \frac{d^dp}{(2\pi)^d}p^4\left( \frac{\partial}{\partial p^2}\frac{1+r_\psi}{p^2(1+r_\psi)^2+k^2\omega} \right)^2\,,\\
m_2^{(F)}(\omega,\eta_\psi)&=-\frac{1}{4v_d}k^{6-d}\tilde\partial_t\int \frac{d^dp}{(2\pi)^d}p^2\left( \frac{\partial}{\partial p^2}\frac{1}{p^2(1+r_\psi)^2+k^2\omega} \right)^2\,,\\
m_{12}^{(FB)}(\omega_\psi,\omega_\phi;\eta_\psi,\eta_\phi)&=-\frac{1}{4v_d}k^{4-d}\tilde\partial_t\int \frac{d^dp}{(2\pi)^d}p^2\frac{1+r_\psi}{p^2(1+r_\psi)^2+k^2\omega_\psi}\frac{\partial}{\partial p^2}\frac{1}{p^2(1+r_\phi)+k^2\omega_\phi}\,.
\end{align}
where we used regulators of the form $R_\phi=Z_\phi p^2r_\phi$ and $R_\psi=Z_\psi i\gamma_\mu p_\mu r_\psi$ and we defined
\begin{align}\label{eq:scalederiv}
\tilde\partial_t=\sum_{\varphi\in\{\phi,\psi\}}\int dp 2p \frac{1}{Z_\varphi}\partial_t (Z_\varphi r_\varphi(p))\frac{\delta}{\delta r_\varphi(p)}\,.
\end{align}

To calculate the functional renormalization group flow equations for the running couplings/functionals, we further have to specify the regulator functions $r_\psi$ and $r_\phi$.
Here, we choose linear regulators, which allow for a fully analytical calculation of the beta functions, i.e.
\begin{align}\label{eq:cutlin}
r_{\psi,k}(q)&=\left(\frac{k}{q}-1\right)\Theta(k^2-q^2)\,,\\
r_{\phi,k}(q)&=\left(\frac{k^2}{q^2}-1\right)\Theta(k^2-q^2)\,,
\end{align}
with the Heavyside step function $\Theta(x)$.
Furthermore they optimize the convergence to the physical solution \cite{litim1,litim2,litim3,litim4}.
Then the threshold functions read
\begin{align}
l_0^{(B)}(\omega,\eta_\phi)&=\frac{2}{D}\left(1-\frac{\eta_\phi}{D+2}\right)\frac{1}{1+\omega}\,,\\
l_0^{(F)}(\omega,\eta_\psi)&=\frac{2}{D}\left(1-\frac{\eta_\psi}{D+1}\right)\frac{1}{1+\omega}\,,\\
l_{nm}^{(FB)}(\omega_\psi,\omega_\phi;\eta_\psi,\eta_\phi)&=\frac{2}{D}\left[\left(1-\frac{\eta_\psi}{D+1}\right)\frac{1}{1+\omega_\psi}+\left(1-\frac{\eta_\phi}{D+2}\right)\frac{1}{1+\omega_\phi}\right]\frac{1}{(1+\omega_\psi)^n(1+\omega_\phi)^m}\,,\\
m_4^{(B)}(\omega,\eta_\phi)&=\frac{1}{(1+\omega)^4}\,,\\
m_{22}^{(B)}(\omega_1,\omega_2,\eta_\phi)&=\frac{1}{(1+\omega_1)^2(1+\omega_2)^2}\,,\\
m_4^{(F)}(\omega,\eta_\psi)&=\frac{1}{(1+\omega)^4}+\frac{1-\eta_\psi}{D-2}\frac{1}{(1+\omega)^3}-\left(\frac{1-\eta_\psi}{2D-4}+\frac{1}{4}\right)\frac{1}{(1+\omega)^2}\,,\\
m_2^{(F)}(\omega,\eta_\psi)&=\frac{1}{(1+\omega)^4}\\
m_{12}^{(FB)}(\omega_\psi,\omega_\phi;\eta_\psi,\eta_\phi)&=\left(1-\frac{\eta_\phi}{D+1}\right)\frac{1}{(1+\omega_\psi)(1+\omega_\phi)^2}.
\end{align}
%

\end{widetext}



\end{document}